\documentclass[aps,prd,superscriptaddress,nofootinbib,longbibliography,amsmath,amsfonts,preprintnumbers,longbibliography,notitlepage,10pt,english]{revtex4-1}
\usepackage{amsmath}
\usepackage{amssymb}
\usepackage{amsfonts}
\usepackage{babel}

\makeatletter

\@ifundefined{textcolor}{}
{%
 \definecolor{BLACK}{gray}{0}
 \definecolor{WHITE}{gray}{1}
 \definecolor{RED}{rgb}{1,0,0}
 \definecolor{GREEN}{rgb}{0,1,0}
 \definecolor{BLUE}{rgb}{0,0,1}
 \definecolor{CYAN}{cmyk}{1,0,0,0}
 \definecolor{MAGENTA}{cmyk}{0,1,0,0}
 \definecolor{YELLOW}{cmyk}{0,0,1,0}
 }
\@ifundefined{textcolor}{}{%
 \definecolor{BLACK}{gray}{0}
 \definecolor{WHITE}{gray}{1}
 \definecolor{RED}{rgb}{1,0,0}
 \definecolor{GREEN}{rgb}{0,1,0}
 \definecolor{BLUE}{rgb}{0,0,1}
 \definecolor{CYAN}{cmyk}{1,0,0,0}
 \definecolor{MAGENTA}{cmyk}{0,1,0,0}
 \definecolor{YELLOW}{cmyk}{0,0,1,0}
 }
 
\newcommand{\nn}{\nonumber \\}
\newcommand{\e}{\mathrm{e}}

\usepackage{array,multirow,graphicx}
\usepackage{dcolumn}
\usepackage{txfonts}
\usepackage{newlfont}
\usepackage{times}
\usepackage{bm}
\usepackage[colorlinks,citecolor=blue,urlcolor=blue,linkcolor=blue]{hyperref}
\usepackage[figtopcap]{subfigure}
\usepackage{color}

\makeatother

\begin{document}

\date{\today}

\title{Spherically symmetric black holes with electric and magnetic charge in extended gravity: \\ Physical properties, causal structure, and stability analysis in Einstein's and Jordan's frames}

\author{E. Elizalde}
\email{elizalde@ieec.uab.es}
\affiliation{Institut de Ci\`encies de l'Espai (ICE-CSIC/IEEC),
Campus UAB, c. Can Magrans s/n, 08193, Barcelona, Spain}

\author{G.G.L. Nashed}
\email{nashed@bue.edu.eg}
\affiliation {Centre for Theoretical Physics, The British University, P.O. Box
43, El Sherouk City, Cairo 11837, Egypt}

\author{S. Nojiri}
\email{nojiri@gravity.phys.nagoya-u.ac.jp}
\affiliation{Kobayashi-Maskawa Institute for the Origin of Particles and the Universe, Nagoya University, Nagoya 464-8602, Japan}

\author{S.D. Odintsov}
\email{odintsov@ieec.uab.es}
\affiliation{Institut de Ci\`encies de l'Espai (ICE-CSIC/IEEC),
Campus UAB, c. Can Magrans s/n, 08193, Barcelona, Spain}
\affiliation{Instituci\'o Catalana de Recerca i Estudis Avan\c{c}ats (ICREA),
Barcelona, Spain}
\affiliation{Tomsk State Pedagogical University, 634061 Tomsk, Russia}

\begin{abstract}
Novel static black hole solutions with electric and magnetic charges are derived for  the class of modified gravities: $f({\cal R})={\cal R}+2\beta\sqrt{{\cal R}}$, with or without a cosmological constant. The new black holes behave asymptotically as flat or (A)dS space-times with a dynamical value of the Ricci scalar given by $R=\frac{1}{r^2}$ and $R=\frac{8r^2\Lambda+1}{r^2}$, respectively. They are characterized by three parameters, namely their mass and electric and magnetic charges, and constitute black hole solutions different from those in Einstein's general relativity. Their singularities are studied by obtaining the Kretschmann scalar and  Ricci tensor, which shows a dependence on the parameter $\beta$ that is not permitted to be zero. A conformal transformation is used to display the black holes in Einstein's frame and check if its physical behavior is changed w.r.t. the Jordan one. The thermal stability of the solutions is discussed by using thermodynamical quantities, in particular the entropy, the  Hawking temperature, the quasi-local energy, and the Gibbs free energy. Also, the casual structure of the new black holes is studied, and a stability analysis is performed in both frames using the odd perturbations technique and the study of the geodesic deviation. It is concluded that, generically, there is coincidence of the physical properties of the novel black holes in both frames, although this turns not to be the case for the Hawking temperature.
\end{abstract}


\maketitle
\section{\bf Introduction}
The discovery of gravitational waves (GW) has shed light on a new possibility to probe the laws of physics in strong gravitational fields \cite{Abbott:2016blz}. General relativity (GR) has been confirmed to a very good precision on  weak gravitational field backgrounds \cite{Will:2014kxa}; however, the precise form of the anticipated, necessary modification of GR to deal with strong gravitational fields is not confirmed yet, although different possibilities have been proposed. The discovery of GW definitely provides an excellent chance to test those modified gravity theories in the strong gravitational fields of black hole solutions \cite{TheLIGOScientific:2016src} and neutron stars \cite{TheLIGOScientific:2017qsa}.

The simplest generalizations of GR are the $f({\cal R})$ gravitational theories, whose Lagrangian involves nonlinear
terms in ${\cal R}$.
A simple possibility is power-law gravity, described by a Lagrangian of the form
$$ f({\cal R}) = {\cal R}+\frac{{\cal R}^n}{6m^2},$$ where $n$ is an arbitrary number and with $m^2$ being a positive mass squared. The term  ${\cal R}^2$ has a natural interpretation as corresponding to the lowest order quantum perturbative additions to classical gravity, and it is, at the same time, responsible for the inflation at early epoch. In addition, this term should be seriously  considered when dealing with local objects on the background of a strong gravitational field. related to this, a  numerous of research papers have been focused on the study of  black hole solutions, as e.g.
\cite{Lu:2017kzi,delaCruzDombriz:2009et,Nelson:2010ig,Nojiri:2017kex,Nojiri:2013su,Kehagias:2015ata,Canate:2015dda,Yu:2017uyd,Canate:2017bao,Sultana:2018fkw}, and neutron stars solutions  \cite{Cooney:2009rr,Arapoglu:2010rz,Hanafy:2015yya,Orellana:2013gn,Awad:2017tyz,Astashenok:2013vza,Shirafuji:1997wy,Nashed:2016tbj,Ganguly:2013taa,Nashed:2011fg,
Capozziello:2015yza,Nashed:2018nll,Resco:2016upv,Nashed:2014sea}. We should note that $f({\cal R})$ theories can be related with theories of Brans-Dicke  type (see, e.g., \cite{PhysRev.124.925}), in particular with the ones involving  a scalar and a potential of gravitational origin \cite{PhysRevLett.29.137,Chiba:2003ir}. Similar to what happens with black holes, in Brans-Dicke  theories involving a potential with a squared positive mass, a ``no-hair (B theorem)'' holds, preventing  the appearance  of  non-trivial scalar hair \cite{hawking1972,PhysRevD.51.R6608}. The same
 theorem  forbids  the existence of hairy black hole solutions in the case of  the ${\cal R}^2$ model \cite{delaCruzDombriz:2009et,Nelson:2010ig,Yu:2017uyd,Canate:2017bao}.
Several black holes have been got already for $f({\cal R})$ theories \cite{delaCruzDombriz:2009et,Nashed:2018piz,Moon:2011hq,2018EPJP..133...18N,PhysRevD.94.024062,2018IJMPD..2750074N,Ca_ate_2016,
Moon:2011fw,AyonBeato:2010tm,Hendi:2011eg,
 Hendi:2014mba,2013PhRvD..87j4029C}. And their physical properties are discussed in, e.g., \cite{Addazi:2016hip,PhysRevD.91.064009,Akbar:2006mq,Faraoni:2010yi}.

The observation of the mathematical similarity between gravitational and electromagnetic fields goes back at least to the eighteenth century, when  Coulomb constructed his inverse square law to formulate the force between two charges at a distance $r$ \cite{Griffiths:1492149}. Coulomb's law is, in this sense, a complete analogue of the gravitational law \cite{Newton:1687eqk} for the force acting on two masses separated  by the same distance. The similarity between this expression for the two forces led scientists  to conjecture that the gravitational force  exerted by the sun on the planets could be accompanied by a  magnetic force leading to the precession of their orbits and, thence, they would investigate from this standpoint the discrepancy found by Newton in the precession of  Mercury's orbit. In fact, Mercury's perihelion precession was definitely explained by Einstein's GR, sometime after this similarity between gravitational and  electromagnetic fields had been exploited, in
some regimes. Moreover, it is known that gravitation involves a gravitomagnetic field because of the mass current \cite{1918PhyZ...19...33T,1918PhyZ...19..156L,lense1918influence,10.2307/j.ctv301gk5}. Additionally, Einstein GR forecasts  a gravitomagnetic field because of the proper rotation of the Sun that effects the planetary orbits \cite{Mashhoon1984,1916MNRAS..77..155D,Dass:2019kon}. Those are well-known facts. The aim of the present paper is to construct brand new black hole solutions\footnote{The form of $f(R)$  presented in this study is different from the one in \cite{Nashed:2019tuk}. Also the forms of the black holes derived  here are different from \cite{Nashed:2019tuk} because the charge term in this study does not depend on the parameter of the higher order curvature while it depends in \cite{Nashed:2019tuk}.}, possessing electric and magnetic charge, within the family of $f({\cal R})$ modified gravities, to describe them in both the Jordan and the Einstein frames, and to study a number of their physical properties, by calculating associated thermodynamical quantities. Moreover, we will study their causal structure and perform a detailed stability analysis by using odd perturbation techniques and the study of the geodesic deviation.

The formulation of this study is  as follows.  A brief introduction to  the theory of  Maxwell-$f({\cal R})$ gravity is given in Sec. \ref{S2}. In Sec. \ref{S3}, restricting to spherical symmetry, an exact solution of the field equations of the Maxwell-$f({\cal R})$ theory is obtained. In Sec. \ref{S4}, the same derivation is performed for the case of the Maxwell-$f({\cal R})$ theory  involving a cosmological constant. We obtain a novel solution which  asymptotically has AdS or dS space. We discuss in detail the characteristic features of these solutions in Sec. \ref{S55}. In Sec. \ref{S333}, by  using conformal transformation, we derive the black hole solutions in the Einstein frame. In Sec. \ref{S6666}, basic  thermodynamical quantities, such as the entropy, quasi-local energy, the Hawking temperature, and the Gibbs energy are calculated in both the Einstein and the Jordan frames. These calculations show that (with the sole exception of the Hawking temperature) the physical behavior of the black holes obtained do not change generically in going from one to the other frame.  Using the odd perturbations technique, the analysis of the linear stability of the solutions obtained in Secs. \ref{S3}, \ref{S4} and \ref{S333} is performed in Sec.  \ref{S626}. Stability conditions, in relation with  geodesic motion are obtained in Sec. \ref{S9}. And the causal structure of the solution derived in Sec. \ref{S3} is discussed in Sec. \ref{S66666}.  Finally, in Sec. \ref{S77} we present a summary of the main results of this work, draw some compelling conclusions, and discuss some ideas for future work.

\section{Brief note on the Maxwell--$f({\cal R})$ theory}\label{S2}
The   theory of $f({\cal R})$ gravity is an extension of Einstein's GR, first discussed  in \cite{1970MNRAS.150....1B,Capozziello:2011et,Nojiri:2010wj,Nojiri:2017ncd,Capozziello:2003gx,Capozziello:2002rd,Nojiri:2003ft,Carroll:2003wy}. The Lagrangian of this theory is
\begin{eqnarray} \label{a1}
{\mathop{\mathcal{ L}}}:=
{\mathop{\mathcal{ L}}}_g+{\mathop{\mathcal{ L}}}_{E.M.},
\end{eqnarray}
its gravitational term being ${\mathop{\mathcal{ L}}}_g$,  which is given by
\begin{eqnarray} \label{a2} {\mathop{\mathcal{ L}}}_g:=\frac{1}{2\kappa} \int d^4x \sqrt{-g} (f({\cal R})-\Lambda).\end{eqnarray}
 Here ${\cal R}$ represents the Ricci scalar, $\kappa$ is the Newtonian constant, $\Lambda$ is the cosmological constant, $g$ the metric determinant, and $f({\cal R})$ a certain analytic function. Here, we have defined the energy-momentum as ${\mathop{\mathcal{L}}}_{_{_{ E.M.}}} $,  the Lagrangian of the electromagnetic  field, which is given by
 \begin{eqnarray}\label{a3} {\mathop{\mathcal{L}}}_{_{_{
E.M.}}}:=-\frac{1}{2}F^{2},  \end{eqnarray}
where  $F^2=F_{\mu \nu}F^{\mu\nu}$ and $F_{\mu \nu} =2\xi_{[\mu, \nu]}$,  where $\xi_\mu$ is  the gauge potential and comma refers to the ordinary differentiation\footnote{The square brackets stand for anti-symmetrization, i.e. $\xi_{[\mu, \nu]}=\frac{1}{2}(\xi_{\mu, \nu}-\xi_{\nu ,\mu})$ and the rounded ones for symmetrization $\xi_{(\mu, \nu)}=\frac{1}{2}(\xi_{\mu, \nu}+\xi_{\nu ,\mu})$.} \cite{Hendi:2012nj}.

 Performing the variations of the Lagrangian of Eq. (\ref{a1}) w.r.t. the metric tensor $g_{\mu \nu}$ and w.r.t. the strength tensor $F$, respectively, one gets the  field equations of the Maxwell-$f({\cal R})$ theory, in the form \cite{2005JCAP...02..010C}
\begin{eqnarray} \label{f1}
\zeta_{\mu \nu}={\cal R}_{\mu \nu} f_{\cal R}-\frac{1}{2}g_{\mu \nu}f({\cal R})-2g_{\mu \nu}\Lambda +g_{\mu \nu} \Box f_{\cal R}-\nabla_\mu \nabla_\nu f_{\cal R}-8\pi T_{\mu \nu}\equiv0,\end{eqnarray}
\begin{equation}\label{fe2}
\partial_\nu \left( \sqrt{-g} {\textrm F}^{\mu \nu} \right)=0, \end{equation}
where ${\cal R}_{\mu \nu}$ is the Ricci tensor \footnote{The Ricci tensor is defined as
 \[{\cal R}_{\mu \nu}={\cal R}^{\rho}{}_{\mu \rho \nu}=  2\Gamma^\rho{}_{\mu [\nu,\rho]}+2\Gamma^\rho{}_{\beta [\rho}\Gamma^\beta{}_{\nu] \mu},\]
 where $\Gamma^\rho{}_{\mu \nu}$ are the Christoffel symbols of the second kind.}
and $\Box$ is the d'Alembertian operator, $\nabla_\alpha A^\beta$ is the covariant derivative of the vector $A^\beta$, and ${\displaystyle f_{\cal R}=\frac{df({\cal R})}{d{\cal R}}}$.   Here, we define the energy-momentum tensor, $T_{\mu \nu}$,   as
 \begin{eqnarray} T_{\mu \nu}:=\frac{1}{4\pi}\left({ \textrm g}_{\rho
\sigma}{{\textrm  F}_\nu{}^\rho}{{{\textrm F}}_\mu}^{\sigma}-\displaystyle{1 \over 4}  {\textrm g}_{\mu \nu} F^{2}\right).\end{eqnarray}
Taking the trace in  Eq.~(\ref{f1}), one gets
\begin{eqnarray} \label{f3}
\zeta={\cal R}f_{\cal R}-2f({\cal R})-8\Lambda+3\Box f_{\cal R}=0 \,.\end{eqnarray}
Now, we shall take a specific form for the field Eqs. (\ref{f1}), both with and without a cosmological constant to derive exact charged black holes, which behave asymptotically  as   AdS/dS or flat space-times, respectively.

\section{Black hole solutions with magnetic and electric charge }\label{S3}
In this section, we are going to derive a charged black hole for the following specific model  \begin{equation}\label{fR1} f({\cal R})={\cal R}+2\beta\sqrt{{\cal R}}.\end{equation}\label{f2} To achieve this, we introduce the following spherically symmetric ansatz\footnote{We use this ansatz (\ref{met})  to  find an exact solution. Changing the  ansatz (\ref{met}) will produce complicated field equations  that are not easy to solve.}
\begin{eqnarray} \label{met}
& &  ds^2=-w(r)dt^2+\frac{dr^2}{w(r)}+r^2(d\theta^2+\sin^2d\phi^2).  \end{eqnarray}
 The Ricci scalar of the line-element (\ref{met}) is given by

 \begin{eqnarray}\label{r1}
& & {\cal R}=\frac{2-r^2w''-4rw'-2w}{r^2},
\end{eqnarray}
where $w\equiv w(r)$,  $w'\equiv \frac{dw(r)}{dr}$, and $w''\equiv \frac{d^2w(r)}{dr^2}$.
Using Eqs. (\ref{met}) in (\ref{f1}), (\ref{fe2}) and (\ref{f3}), after using Eq. (\ref{r1}) we get a system of fourth order differential equations which are listed in Appendix A.
The off-diagonal components  of these system,  $( A\cdot 2)$,  $(A\cdot 4)$ and  $( A\cdot 5)$, can be solved to determine the unknown functions  $n$, $p$, $s$, and $k$. Substituting the  values of these function into the diagonal components, as well as into the trace field equation, we obtain
\begin{eqnarray} \label{sol}
& &  w(r)=\frac{1}{2}+\frac{c_1}{ r}+\frac{{q_{_E}}{}^2+{q_{_M}}{}^2}{ r^2}, \qquad q=\frac{{q_{_E}}}{r}, \qquad n=c_2 \theta, \qquad s=c_2r,  \qquad p=c_3 r, \qquad k=c_4 \theta-{q_{_M}}\cos\theta,
\end{eqnarray}
where the $c_i$, $i=1\cdots 4$, are constant, and $q_{_E}$ and $q_{_M}$ are other constants related to the electric and magnetic charge, respectively.
The analytic solution (\ref{sol}) satisfies the system of differential equations presented in Appendix A, including the trace of the field equations, provided that $c_1=\frac{1}{3\beta}$. The Ricci scalar is obtained, in the form
\begin{equation} \label{ri}
{\cal R}=\frac{1}{r^2}\, .\end{equation}
This is a consistency check of the procedure, too. The  solution
(\ref{sol}) metric reads
\begin{eqnarray} \label{met5}
& &  ds^2=-\left(\frac{1}{2}+\frac{1}{3\beta r}+\frac{{\cal K}^2}{r^2}\right)dt^2+\left(\frac{1}{2}+\frac{1}{3\beta r}+\frac{{\cal K}^2}{ r^2}\right)^{-1}dr^2+r^2d\Omega^2,
\end{eqnarray}
where we have set ${q_{_E}}{}^2+{q_{_M}}{}^2={\cal K}^2$. Eq.~(\ref{met5}) behaves  asymptotically  as a flat space-time. Solution (\ref{sol}) coincides  with that obtained in \cite{Sebastiani:2010kv}  when ${\cal K}^2=0$, i.e. $q_E=q_M=0$. Also, the solution obtained (\ref{met5})  corresponds to the spherically symmetric space-time in $f({\cal R})$ gravitational theories, and differs from the corresponding one  in \cite{Nashed:2019tuk} by the more general expression of the 1-form gauge potential, given in Eq. $(A\cdot 9)$, and by the parameter ${\cal K}$  that couples to electric and magnetic fields.

\section{Analytic AdS/dS charged solutions  }\label{S4}
In order to obtain  a black hole solution with charge, behaving asymptotically as AdS/dS,  we take $f(R)$ of the form\footnote{We define ${\textrm R}={\cal R}-8\Lambda$.}
\begin{equation} \label{fR2}
f({\cal R})={\cal R}+2\beta\sqrt{{\cal R}-8\Lambda}.\end{equation}  Using now the anzatz (\ref{met}) in the field Eqs. (\ref{f1}), (\ref{fe2}), and (\ref{f3}), and after applying (\ref{r1}), we obtain a system of fourth order differential equations listed in Appendix B.

Using  the previous procedure, namely solving the off-diagonal components and substituting their values in the diagonal ones,
 we get the  following exact solution
\begin{eqnarray} \label{sol1}
 w(r)=\frac{1}{2}-\frac{2r^2\Lambda}{3}+\frac{1}{3\beta r}+\frac{{q_{_E}}{}^2+{q_{_M}}{}^2}{ r^2}, \qquad q=\frac{{q_{_E}}}{r}, \qquad n=c_2 \theta, \qquad s=c_2r,  \qquad p=c_3 r, \qquad k=c_4 \theta-{q_{_M}}\cos\theta.
\end{eqnarray}
Introducing Eq. (\ref{sol1}) into (\ref{r1}), we obtain the Ricci scalar, as follows
\begin{equation} \label{ri}
{\cal R}=\frac{8r^2\Lambda+1}{r^2}.\end{equation}
The  solution (\ref{sol1}) metric reads
\begin{eqnarray} \label{met3}
& &  ds^2=-\left(\frac{1}{2}-\frac{2r^2\Lambda}{3}+\frac{1}{3\beta r}+\frac{{q_{_E}}{}^2+{q_{_M}}{}^2}{ r^2}\right)dt^2+\left(\frac{1}{2}-\frac{2r^2\Lambda}{3}+\frac{1}{3\beta r}+\frac{{q_{_E}}{}^2+{q_{_M}}{}^2}{ r^2}\right)^{-1}dr^2+r^2d\Omega^2\, ,
\end{eqnarray}
and behaves asymptotically as  AdS/dS space-time. The solution (\ref{sol1}) is different from the one derived in \cite{Sebastiani:2010kv}, owing to the same reason already discussed for the solution (\ref{sol}).

\section{Properties of the found black holes}\label{S55}

For the solution (\ref{sol}), the metric can be put as
\begin{equation} \label{me}
ds^2=-\left(\frac{1}{2}-\frac{m}{ r}+\frac{{\cal K}^2}{ r^2}\right)dt^2+\left(\frac{1}{2}-\frac{m}{ r}+\frac{{\cal K}^2}{ r^2}\right)^{-1}dr^2+r^2d\Omega^2\;, \qquad \textrm{where} \qquad m=-c_1=-\frac{1}{3\beta}.\end{equation}
Eq.~(\ref{me}) indicates that the dimensional parameter $\beta$ cannot vanish. And this says that the line-element is the same as for Reissner-Nordstr\"om space-time when ${\cal K}=q_{_E}$ and $q_{_M}=0$.

The line-element  of the solution (\ref{sol1})  can be written as
\begin{equation} \label{me1}
ds^2=\left(\frac{1}{2}-\frac{2r^2\Lambda}{3}-\frac{m}{r}+\frac{{\cal K}^2}{ r^2}\right)dt^2-\left(\frac{1}{2}-\frac{2r^2\Lambda}{3}-\frac{m}{r}+\frac{{\cal K}^2}{ r^2}\right)^{-1}dr^2-r^2d\Omega^2\;, \qquad \textrm{where, again} \qquad m=-c_1=-\frac{1}{3\beta},\end{equation}
which tells us  that the line-element (\ref{me1}) does behave asymptotically as  AdS/dS and that it coincides with the Reissner-Nordstr\"om space-time when ${\cal K}=q_{_E}$ and $q_{_M}=0$.   Eqs. (\ref{me}) and (\ref{me1}) insure that  $\beta\neq 0$.

We now turn to the regularity of the solutions (\ref{sol}) and (\ref{sol1}), for $w(r)=0$. For  (\ref{sol}), we calculate the scalar invariants, with the result
\begin{eqnarray} \label{scal1}
&&{\cal R}^{\mu \nu \lambda \rho}{\cal R}_{\mu \nu \lambda \rho}= \frac{4r^2-4\beta r^3+3r^4\beta^2+{\cal K}^2[48r \beta-12r^2\beta^2+168\beta^2]+336\beta q_{_M}^2q_{_M}^2}{3\beta^2r^8}, \nonumber\\
& &{\cal R}^{\mu \nu}{\cal R}_{\mu \nu}=\frac{r^4+{\cal K}^2[4r^2+8{\cal K}^2]}{2r^8}, \qquad \qquad {\cal R}= \frac{1}{r^2},
\end{eqnarray}
where  ${\cal R}$, ${\cal R}^{\mu \nu}{\cal R}_{\mu \nu }$, and ${\cal R}^{\mu \nu \lambda \rho}{\cal R}_{\mu \nu \lambda \rho}$ are the Ricci scalar, the Ricci tensor  square, and the Kretschmann scalars, respectively.   Eqs. (\ref{scal1}) show that, at $r=0$, the solutions develop a true singularity in which the dimensional parameter $\beta$  cannot vanish, so that the solution
(\ref{sol}) is not reducible to one of GR. Thus, this black hole  is a brand new one of the class of $f({\cal R})$ modified theories.

Employing Eq. (\ref{sol1}),  we obtain the scalar invariants, as
\begin{eqnarray} &&{\cal R}^{\mu \nu \lambda \rho}{\cal R}_{\mu \nu \lambda \rho} =\frac{4r^2-4\beta r^3+3r^4\beta^2+{\cal K}^2[48r \beta-12r^2\beta^2+168\beta^2]+336\beta q_{_M}^2q_{_M}^2+8r^6\beta^2\Lambda[4r^2\Lambda +1]}{3\beta^2r^8},
 \nonumber\\
&&{\cal R}^{\mu \nu}{\cal R}_{\mu \nu} =\frac{r^4+{\cal K}^2[4r^2+8{\cal K}^2]+8r^6\beta^2\Lambda[4r^2\Lambda +1]}{2r^8},\qquad \qquad \qquad
{\cal R} =\frac{8r^2\Lambda+1}{r^2}.\nonumber\\
\end{eqnarray}
The same considerations already done for the solution (\ref{sol}) can also be applied now to the solution (\ref{sol1}), what will insure also that  (\ref{sol1}) is a brand new, charged solution constructed in the class of $f({\cal R})$ gravities, and that it cannot possibly be reduced to a GR solution.
\section{Charged black hole solutions in the Einstein frame}\label{S333}
In this section we will construct charged black holes in the Einstein frame. We thus start  with a brief description of $f(R)$ theories in the Einstein frame.  It is rather well-know that  $f(R)$ gravitational theories can be rewritten under the form of a Brans-Dicke theory, by involving a subsidiary field, $\psi$, through a non-minimal coupling term, as
\begin{equation}\label{J-action2}
{\cal L}:= \int d^4x \sqrt{-g}\, \left[\frac{1}{2\kappa} f_\psi(\psi)(R-\Lambda)-\left(\frac{\psi f_\psi(\psi)-f(\psi)}{2\kappa}\right)\right]+{\mathop{\mathcal{ L}}}_{_{E.M.}},
\end{equation}
where $f_\psi(\psi)=\frac{f(\psi)}{d \psi}$ and ${\mathop{\mathcal{ L}}}_{_{E.M.}}$ is the Lagrangian of the electromagnetic field, given by Eq. (\ref{a3}).
Variation of Eq. (\ref{J-action2}) w.r.t. $\psi$ gives $f_{\psi\psi}(R-\Lambda-\psi)=0$. For $f_{\psi\psi}\neq 0$, one can obtain $\psi=R-\Lambda$ and the above action returns back to the one of  Eq. (\ref{a1}). This means that the field equations produced by the action (\ref{J-action2}) exactly coincide with those previously derived from the Lagrangian  (\ref{a1}), namely (\ref{f1}) and (\ref{fe2}).

When choosing $\sigma=f_{\psi}(\psi)$, the Lagrangian (\ref{J-action2}) is termed as a Brans--Dicke's like theory with a non-minimal coupling term $\sigma R$ and a scalaron potential $V(\sigma)$.  It is well-know that the non-minimal coupling term  can be eliminated from the Jordan frame, by moving to the \textit{Einstein} frame, using the following conformal transformation
\begin{eqnarray} \label{conf-trans}
g_{\mu \nu} \to  {\bar g}_{\mu \nu}(x)=\Omega^2(x) g_{\mu \nu}(x),
\end{eqnarray}
where the space-time conformal factor has been chosen as $\Omega^2(x) =f_{\cal R}$, what demands that $f_{\cal R} > 0$ \cite{Bahamonde:2017kbs,Bahamonde:2016wmz}. From the transformation (\ref{conf-trans}), one can show that the Ricci scalar transforms as $R\to \bar{R}$.  Using now the canonical scalar field
\begin{equation} \psi=\sqrt{\frac{6}{\kappa}}\, \ln \Omega =\sqrt{\frac{3}{2\kappa}}\, \ln f_{\cal R},\end{equation}
and from the conformal transformation (\ref{conf-trans}), the Lagrangian (\ref{J-action2}) converts into a scalar-tensor theory in the Einstein frame,  as 
\begin{eqnarray} \label{E-action1}
{\mathop{\mathcal{ L}_E}}:= \int d^4x \sqrt{-{\bar g}} \left[\frac{1}{2\kappa}({\bar{\cal  R}}-\Lambda)-\frac{1}{2}{{\bar g}}^{\mu \nu}\partial_\mu \psi \partial_\nu \psi-V(\psi)\right]+\mathcal{L}_{_{E.M.}},
\end{eqnarray}
where
\begin{eqnarray} \label{pot}
V_E(\psi)=\displaystyle\frac{{\cal R}f_{\cal R}-f}{2\kappa {f_{\cal R}}^2}=\frac{V(\psi)}{f_R^2},\end{eqnarray}
 is the potential of the canonical scalar field $\psi$. The potential $V_E(\psi)$ can be rewritten in terms of $\psi$ by using the inverse relation $f_{\cal R}=e^{ \sqrt{2\kappa/3}\,\psi}$. Performing the conformal transformation (\ref{conf-trans}), the energy--momentum tensor converts into \cite{Hendi:2009sw,Bhattacharya:2017pqc,Capozziello:2010sc,Bahamonde:2016wmz,Bamba:2015uma,Chakraborty:2018ost}
\begin{equation} T_{\mu\nu} \to \bar{T}_{\mu\nu}=\Omega(x)^{-2} T_{\mu\nu}.\end{equation}

\begin{figure}
\centering
 \includegraphics[scale=0.4]{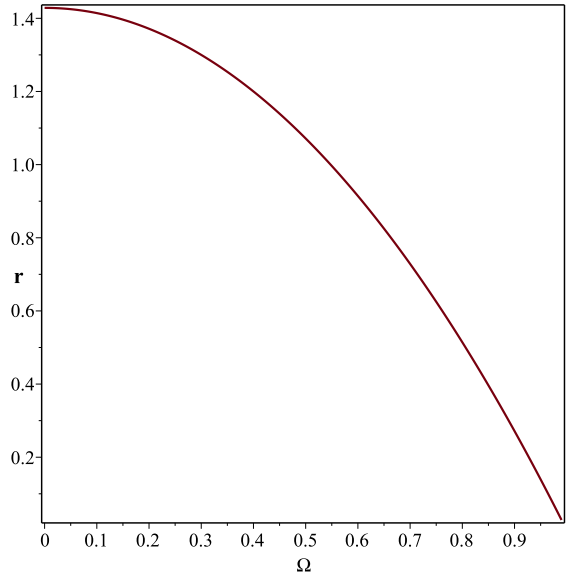}\hspace{0.2cm}
\caption{Plot of the radial coordinate, $r$,  in terms of the scalar field,  $\Omega$}
\label{Fig:1}
\end{figure}

\begin{figure}
\centering
 \includegraphics[scale=0.4]{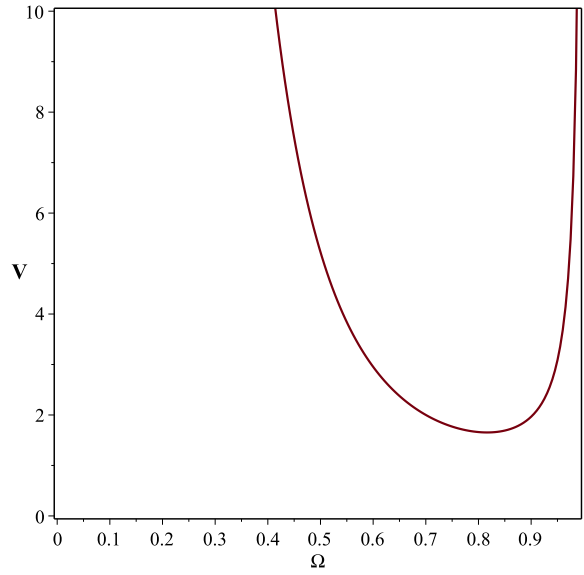}\hspace{0.2cm}
\caption{Schematic plot of the scalar field  $\Omega$  versus the potential  $V$.}
\label{Fig:2}
\end{figure}

In the following,  we are going to apply the conformal transformation (\ref{conf-trans}) to the space-time metric (\ref{me}), i.e. $d\bar{s}_E^2=\Omega^2 ds_J^2$, where the conformal factor of the $f(R)$ gravity (\ref{fR1}) is given by
\begin{equation}
\Omega^2=f_R=1+r\beta.
\end{equation}
The relation between the scalar field $\Omega$ and the radial coordinate $r$ is plotted in Fig. \ref{Fig:1}.
Finally, using Eq. (\ref{pot}), the potential of this model reads
\begin{equation}\label{pot1}
V(r)=-\frac{\beta}{2\kappa^2r(1+\beta r)^2},  \Rightarrow V(\Omega)=\frac{\beta^2}{2\kappa^2\Omega^4(1-\Omega^2)}.
\end{equation}
Eq. (\ref{pot1}) is plotted in Fig.\ref{Fig:2}.

Thus, we can write the Einstein frame metric as
\begin{eqnarray}\label{line_element_1E}
\nonumber    d\bar{s}_E^2&=& \Omega^2\left[-w(r) dt^2+\frac{dr^2}{w(r)}+r^2 \left(d\theta^2+\sin^2 \theta d\phi^2\right)\right],\\
    &=&-\bar{w}(\bar{r}) dt^2+\frac{d\bar{r}^2}{\bar{w_1}(\bar{r})}+\bar{r}^2 \left(d\theta^2+\sin^2 \theta d\phi^2\right),
\end{eqnarray}
where
\begin{eqnarray}\label{conf}
&&\bar{r}= \Omega(r)r, \qquad \bar{w}(\bar{r})=w(r(\bar{r}))=\Omega^2(\bar{r})w(\bar{r}), \quad  \bar{w}_1(\bar{r})=w_1(r(\bar{r}))=\frac{\Omega^2(\bar{r})[x^2(\bar{r})+12x(\bar{r})+144]^2}{16x(\bar{r})[x^2(\bar{r})+6x(\bar{r})+144]}\,, \quad \textrm {where} \nonumber\\
&& x(\bar{r})=\Bigg\{12\Big[9\beta \bar{r}+\sqrt{81\beta^2\bar{r}^2-12}\Big]\Bigg\}^{2/3}. \end{eqnarray}

Eq.~(\ref{conf}) shows  that the solutions (\ref{sol}) and (\ref{sol1}) have been deformed due to the conformal transformation (\ref{conf-trans}) and that the dimensional parameter $\beta$ must satisfy $\beta>\frac{2}{3\sqrt{3}\bar{r}}$. Is this deformation effect conveying the physics in both the Jordan and the Einstein frames? We will answer this question in the next section.

\section{Black hole thermodynamics in the Jordan frame}\label{S6666}
The Hawking temperature is usually defined as \cite{PhysRevD.86.024013,Sheykhi:2010zz,Hendi:2010gq,PhysRevD.81.084040}
  \begin{equation}
T_+ = \frac{w'(r_+)}{4\pi},
\end{equation}
where the event horizon $r = r_+$ is the positive solution of the equation  $w(r_+) = 0$ which satisfies $w'(r_+)\neq 0$.
In the framework of $f({\cal R})$ gravity, the entropy   is given by \cite{PhysRevD.84.023515,Zheng:2018fyn}
\begin{equation}\label{ent}
S(r_+)=\frac{1}{4}Af_{{\cal R}}(r_+),
\end{equation}
where $A$  represents the area. The  quasi-local  energy is defined  in the context  of $f({\cal R})$ theory  as \cite{PhysRevD.84.023515,Zheng:2018fyn}
\begin{equation}\label{en}
E(r_+)=\frac{1}{4}\displaystyle{\int }\Bigg[2f_{{\cal R}}(r_+)+r_+{}^2\Big\{f({\cal R}(r_+))-{\cal R}(r_+)f_{{\cal R}}(r_+)\Big\}\Bigg]dr_+.
\end{equation}

\begin{figure}
\centering
 \subfigure[~Spherically symmetric space-time]{\label{fig:33a}\includegraphics[scale=0.4]{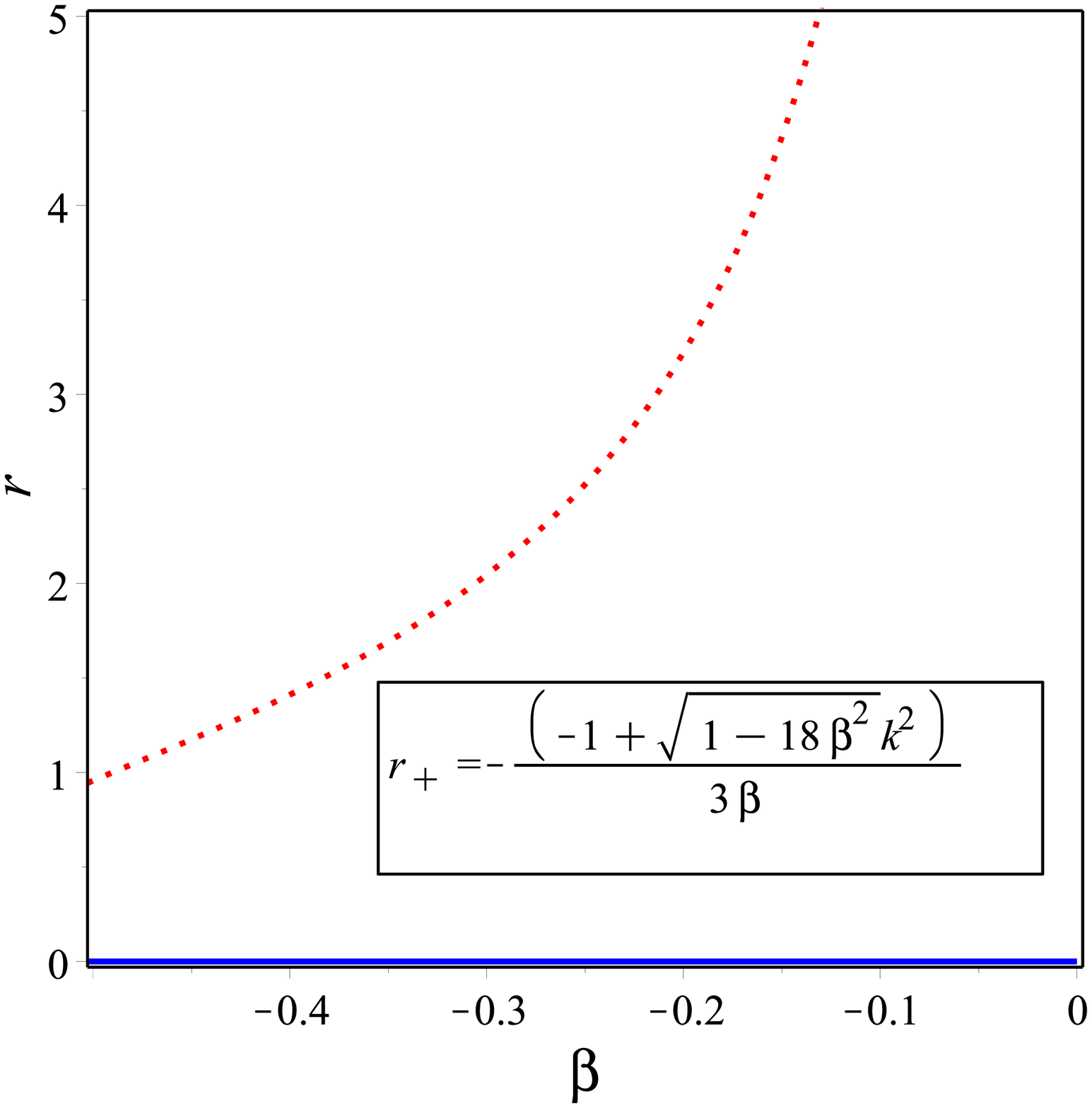}}\hspace{0.2cm}
\subfigure[~Spherically symmetric AdS/dS space-times]{\label{fig:33b}\includegraphics[scale=0.4]{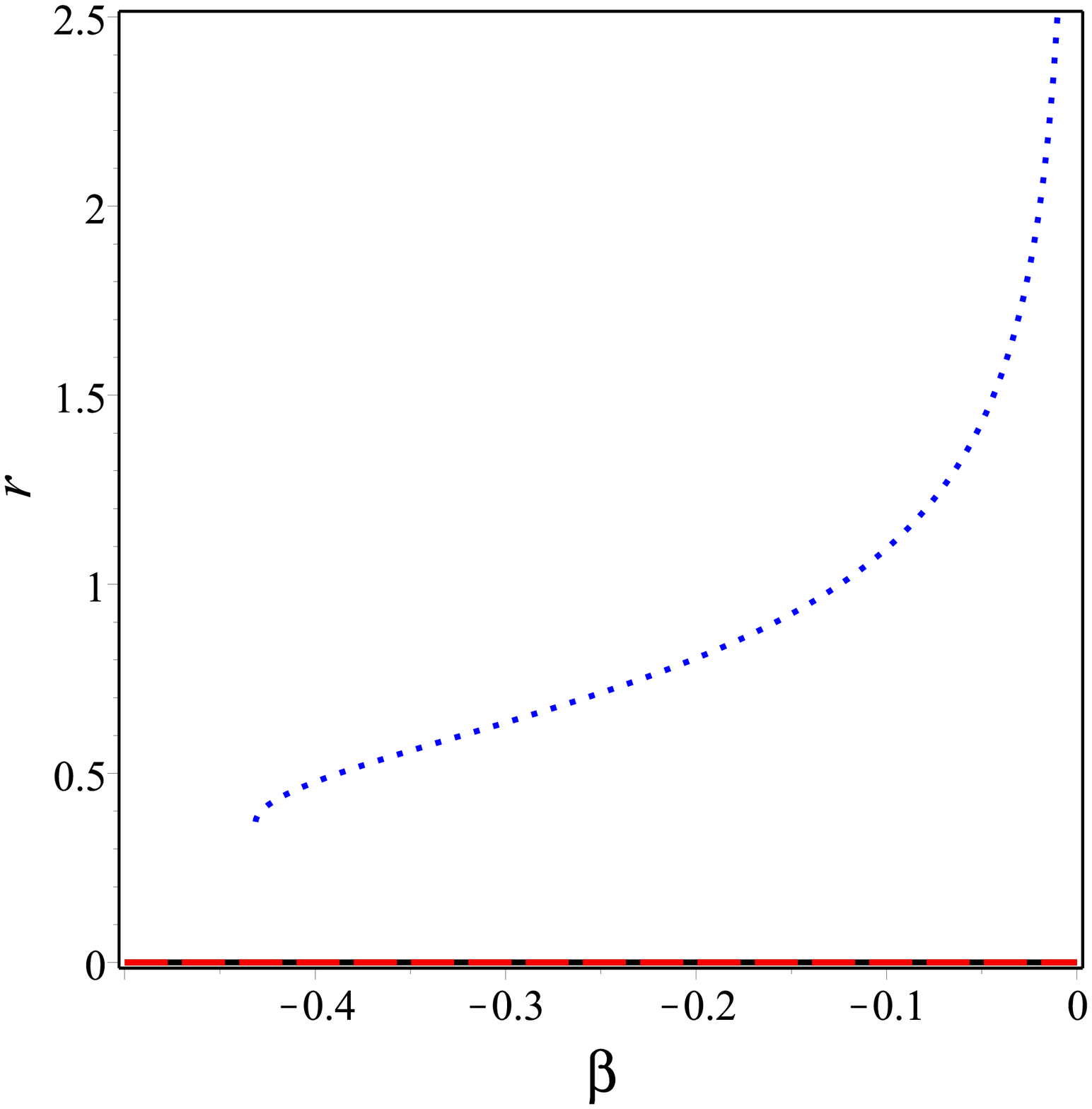}}
\caption{Plot of  $r$  versus $\beta$ for the black holes (\ref{sol}) and (\ref{sol1}), respectively.}
\label{Fig:33}
\end{figure}

The constraint $w(r_+) = 0$ yields
\begin{eqnarray} \label{m33}
&&   {r_+}_{{}_{{}_{{}_{{}_{\tiny Eq. (\ref{sol})}}}}}=-\frac{1}{3\beta}\left[1+\sqrt{1-18\beta^2{\cal K}^2}\right], \qquad \qquad {r_-}_{{}_{{}_{{}_{{}_{\tiny Eq. (\ref{sol})}}}}}=\frac{1}{3\beta}\left[\sqrt{1-18\beta^2{\cal K}^2}-1\right],
 \nonumber\\
&&{r_+}_{{}_{{}_{{}_{{}_{\tiny Eq. (\ref{sol1})}}}}}=Root(4y^4\beta \Lambda+2y-3\beta y^2-6\beta{\cal K}^2 ),
\end{eqnarray}
where $Root(4y^4\beta \Lambda-3\beta y^2+2y+2)$ are the roots of the equation $(4y^4\beta \Lambda+2y-3\beta y^2-6\beta {\cal K}^2=0)$, which is proven to have one  real root.  The first equation of  (\ref{m33})  The first of Eqs. (\ref{m33}) tells us that the  parameter  $\beta$ cannot vanish, therefore the solution (\ref{sol}) has no corresponding one  in the GR limit. Also,  Eq. (\ref{m33})  indicates  that the parameter $\beta$ must be negative, therefore, when there is no charge the horizons have a positive real value. Furthermore, Eq. (\ref{m33}) also sets constraints on $\beta$, namely $\beta<\frac{1}{3{\cal K}\sqrt{2}}$.
 The  behavior of the radial coordinate, $r$, in terms of the parameter $\beta$  is plotted in Fig. \ref{Fig:33}\subref{fig:33a}. Also, we plot there the behavior of the radial coordinate $r$ and the parameter $\beta$ for the third equation of (\ref{m33})\footnote{The values of the electric and the magnetic fields, and of the cosmological constant $\Lambda$ to be used in our discussion are, respectively: ${\cal K}=-0.6, \ \i.e., \ \  q_{_E}=q_{_m}=-0.3, \ \ \Lambda=-3$.  The value of the parameter ${\cal K}$ is consistent with the restriction $\beta<\frac{1}{3{\cal K}\sqrt{2}}$. In Fig. \ref{Fig:33}\subref{fig:33b} the plot is drawn against $r$, which is the positive real root of Eq. $Root(4y^4\beta \Lambda-3\beta y^2+2y+2)$.}. We now resume our consideration of the thermodynamics, assuming $\beta<0$, in accordance with the previous discussion, and considering the outer event  horizon, $r_+$, which agrees with the constraint $\beta<0$.

\begin{figure}
\centering
\subfigure[~Plot of the black hole solution's entropy ]{\label{fig:2a}\includegraphics[scale=0.4]{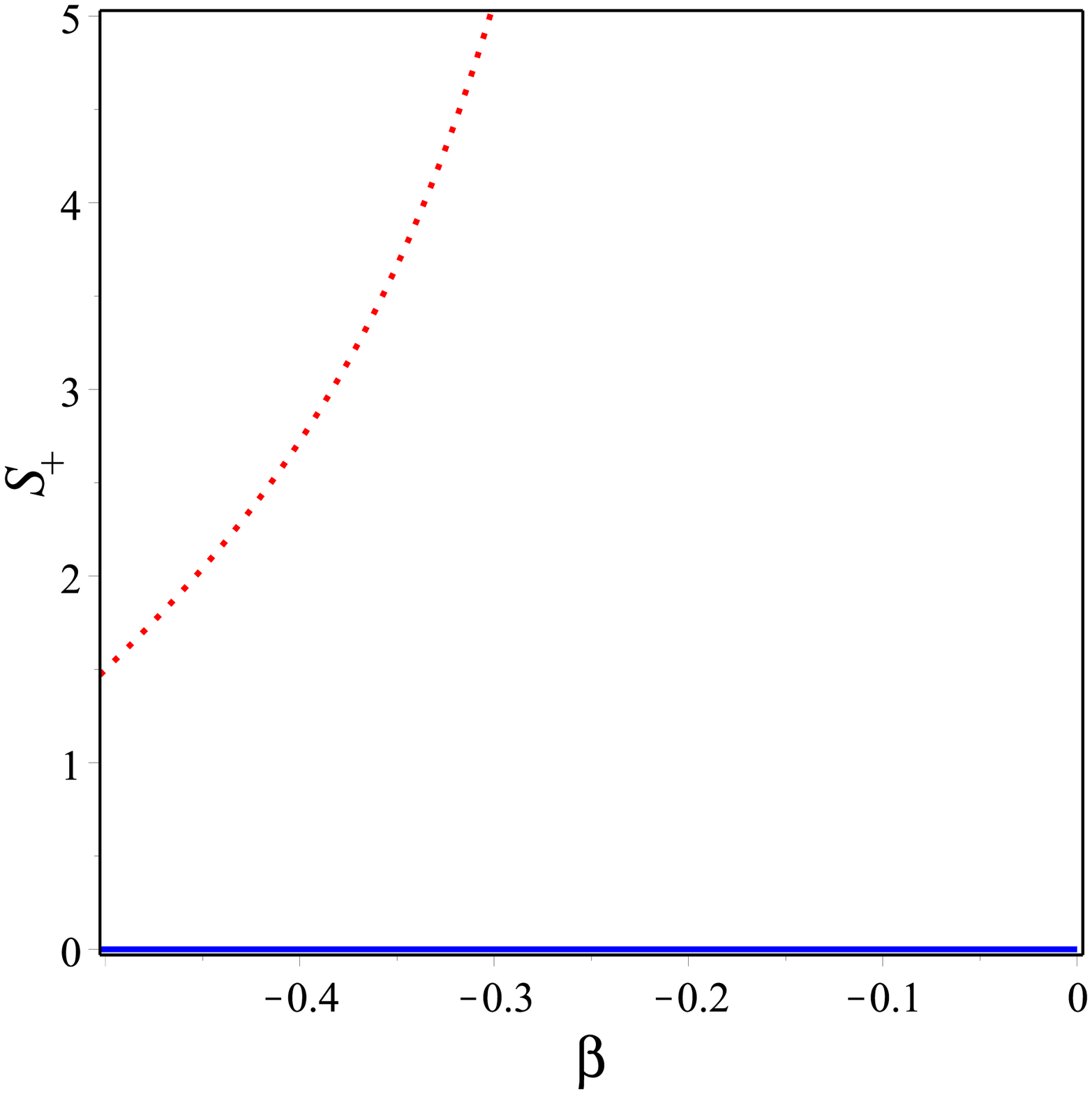}}\hspace{0.2cm}
\subfigure[~Plot of the black hole solution's entropy]{\label{fig:2b}\includegraphics[scale=0.4]{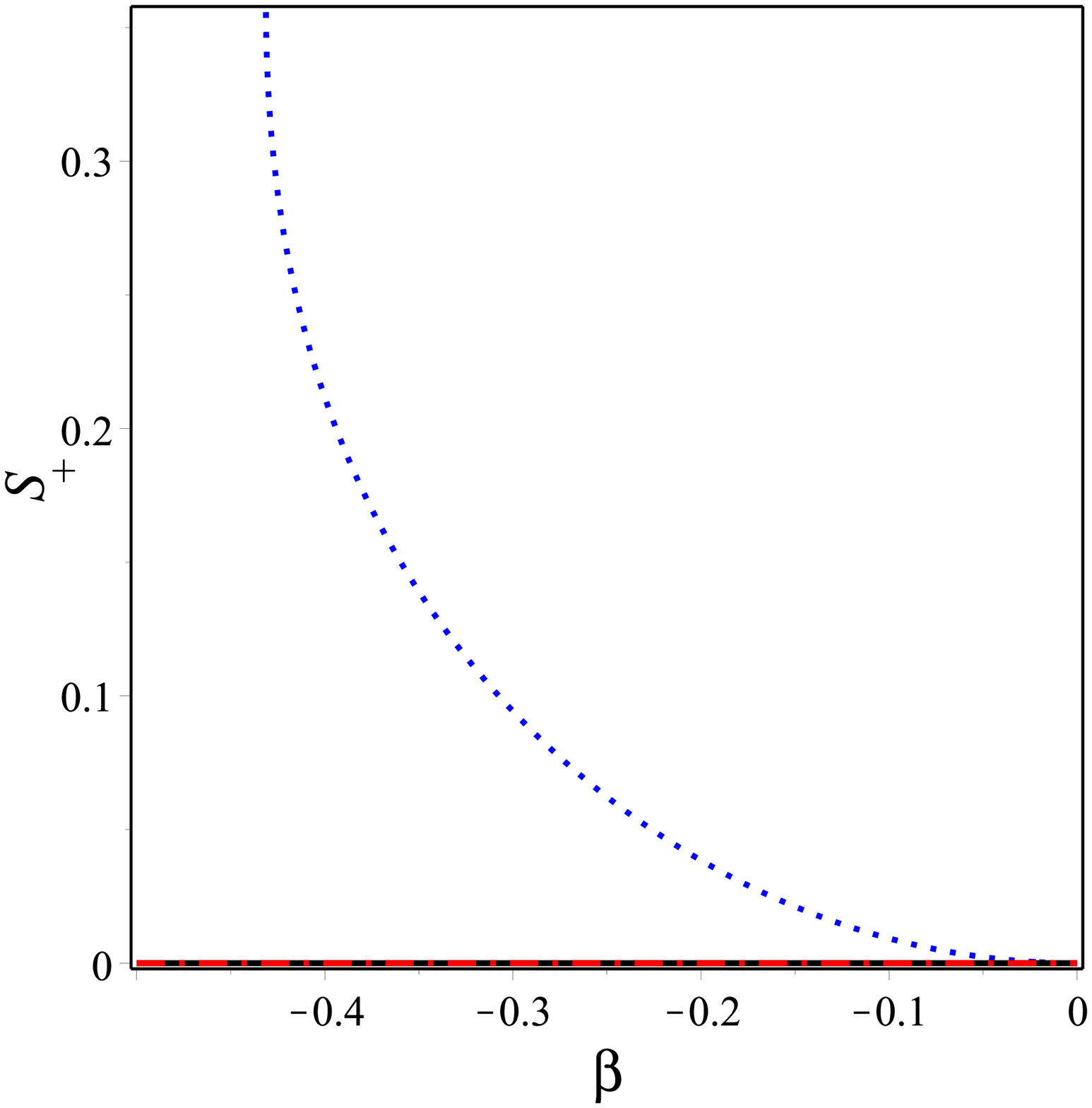}}
\caption{{Plot of the entropy of the solutions (\ref{sol}) and (\ref{sol1}) versus  $\beta$.}}
\label{Fig:44}
\end{figure}

From Eq. (\ref{ent}), the entropy of solutions (\ref{sol}) and (\ref{sol1}) takes the form
\begin{eqnarray} \label{ent1}
{S_+}_{{}_{{}_{{}_{{}_{\tiny Eq. (\ref{sol})}}}}}&=&\frac{\pi}{27\beta^2}\left[1- \sqrt{1-18\beta^2 {\cal K}^2}\right]^2\left[2+\sqrt{1-18\beta^2{\cal K}^2}\right], \nonumber\\
{S_+}_{{}_{{}_{{}_{{}_{\tiny Eq. (\ref{sol1})}}}}}&=&\frac{\pi  r_+{}^2}{4}\left[1+\beta r_+\right].
\end{eqnarray}
 The first  of Eqs. (\ref{ent1}) clearly proves that we have a positive entropy, all the time.  The second  of Eqs. (\ref{ent1}) indicates that $\beta<-\frac{1}{r_+}$, so as to obtain a positive entropy. Eqs.
(\ref{ent1}) are represented in Fig. \ref{Fig:44}.  Observe the entropy $S$  not being proportional to $A$, because of Eq.
(\ref{ent}). And also that $S$ is in fact proportional to $A$ (as it should) provided of course that $f_{\cal R}=1$.

\begin{figure}
\centering
\subfigure[~The black hole solution's temperature in Jordan's frame]{\label{fig:3a}\includegraphics[scale=0.4]{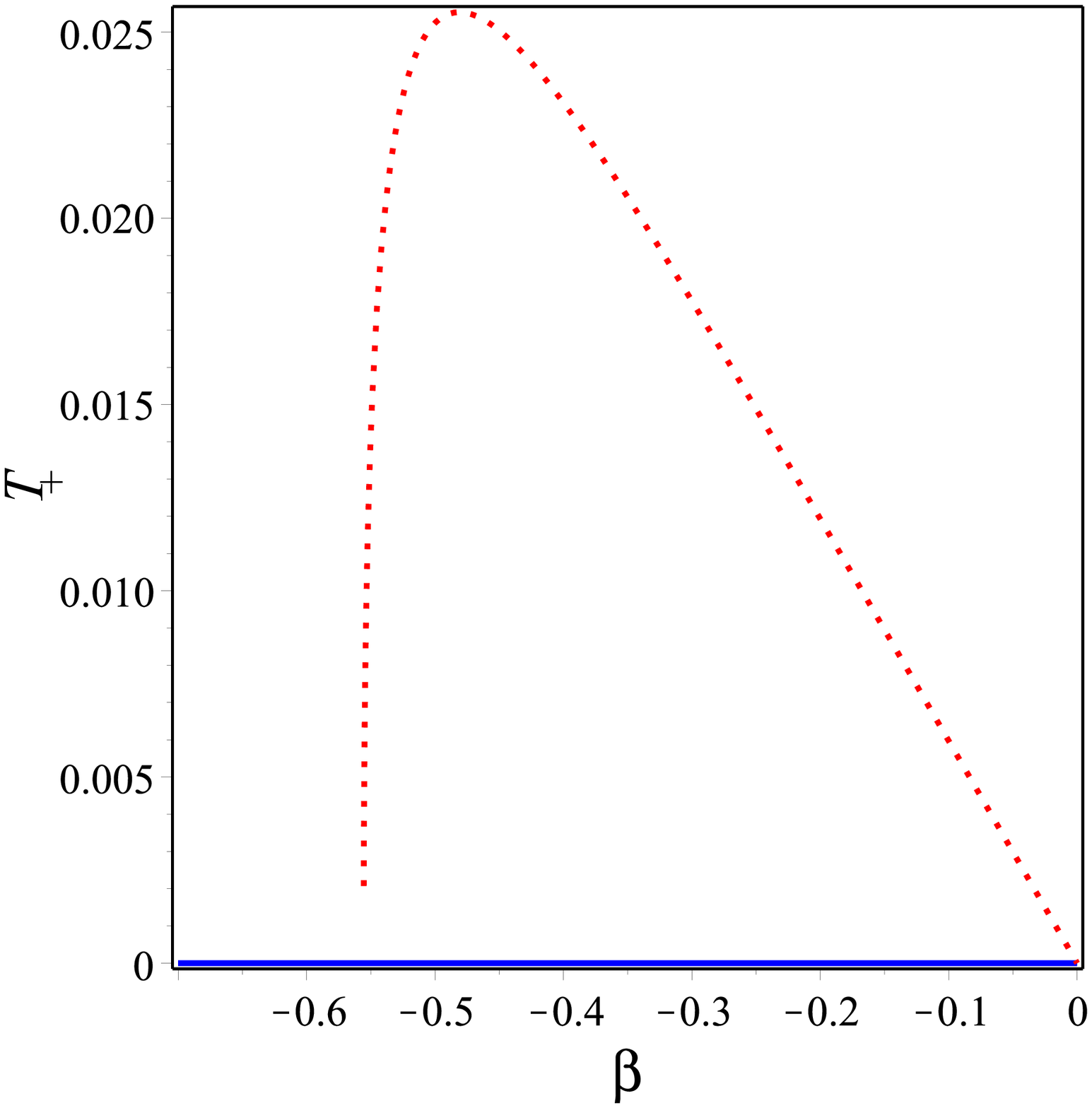}}\hspace{0.2cm}
\subfigure[~The black hole solution's temperature in Jordan's frame]{\label{fig:3b}\includegraphics[scale=0.4]{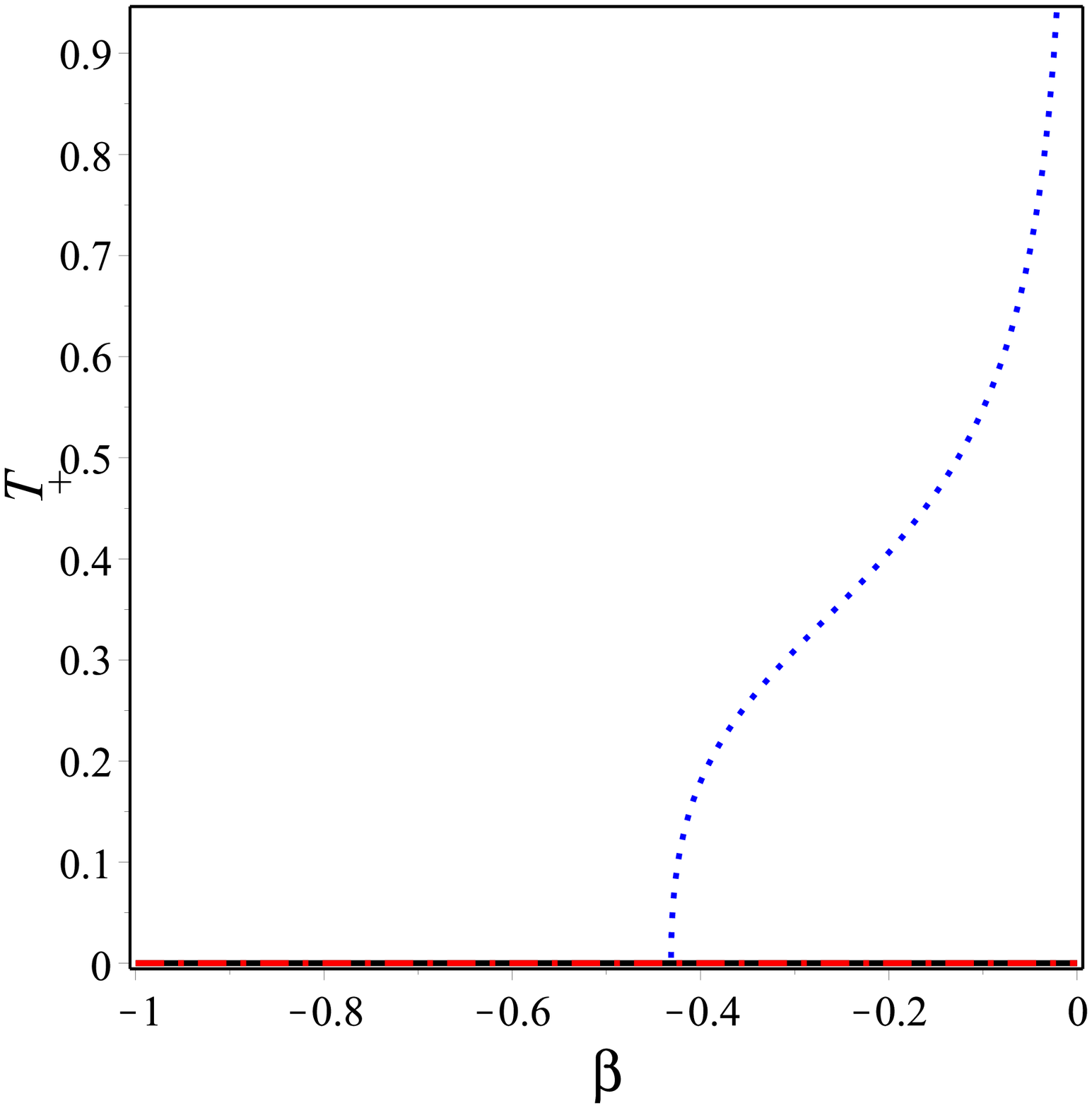}}
\caption{{ Plot of the black hole's temperature in the Jordan frame in terms of the dimensional parameter $\beta$.}}
\label{Fig:3}
\end{figure}

The Hawking temperatures of solutions (\ref{sol}) and (\ref{sol1}) are, respectively,
\begin{eqnarray} \label{m44e}
{T_+}_{{}_{{}_{{}_{{}_{\tiny Eq. (\ref{sol})}}}}}&=&-\frac{3\beta\left[1-18\beta^2 {\cal K}^2-\sqrt{1-18\beta^2 {\cal K}^2}\right]^2}{4\pi(1-\sqrt{1-18\beta^2 {\cal K}^2})^3},\nonumber\\
{T_+}_{{}_{{}_{{}_{{}_{\tiny Eq. (\ref{sol1})}}}}}&=&-\frac{2\beta[ 2\Lambda r_+{}^4+3{\cal K}^2]+r_+}{12\pi \beta r_+{}^3},
\end{eqnarray}
with ${T_+}$ being the  temperature of Hawking's at the event horizon. We  depicted the Hawking's
temperature in Fig. \ref{Fig:3}.   Fig. \ref{Fig:3}
\subref{fig:3a}  proves that we do
have a   positive temperature for the black hole (\ref{sol}) and also Fig.  \ref{Fig:3} \subref{fig:3b} shows that  the temperature  is always positive  for the black hole (\ref{sol1}).

From Eq. (\ref{en}), the quasi-local energy of  (\ref{sol}) and (\ref{sol1}), is obtained as
\begin{eqnarray} \label{m444}
{E_+}_{{}_{{}_{{}_{{}_{\tiny Eq. (\ref{sol})}}}}}&=&-\frac{1+9\beta^2 {\cal K}^2-\sqrt{1-18\beta^2 {\cal K}^2}}{12 \beta},\nonumber\\
{E_+}_{{}_{{}_{{}_{{}_{\tiny Eq. (\ref{sol1})}}}}}&=&\frac{r_+}{8}\left(4+3\beta r_+\right).
\end{eqnarray}
\begin{figure}
\centering
\subfigure[~The black hole's (\ref{sol}) quasilocal energy ]{\label{fig:4a}\includegraphics[scale=0.4]{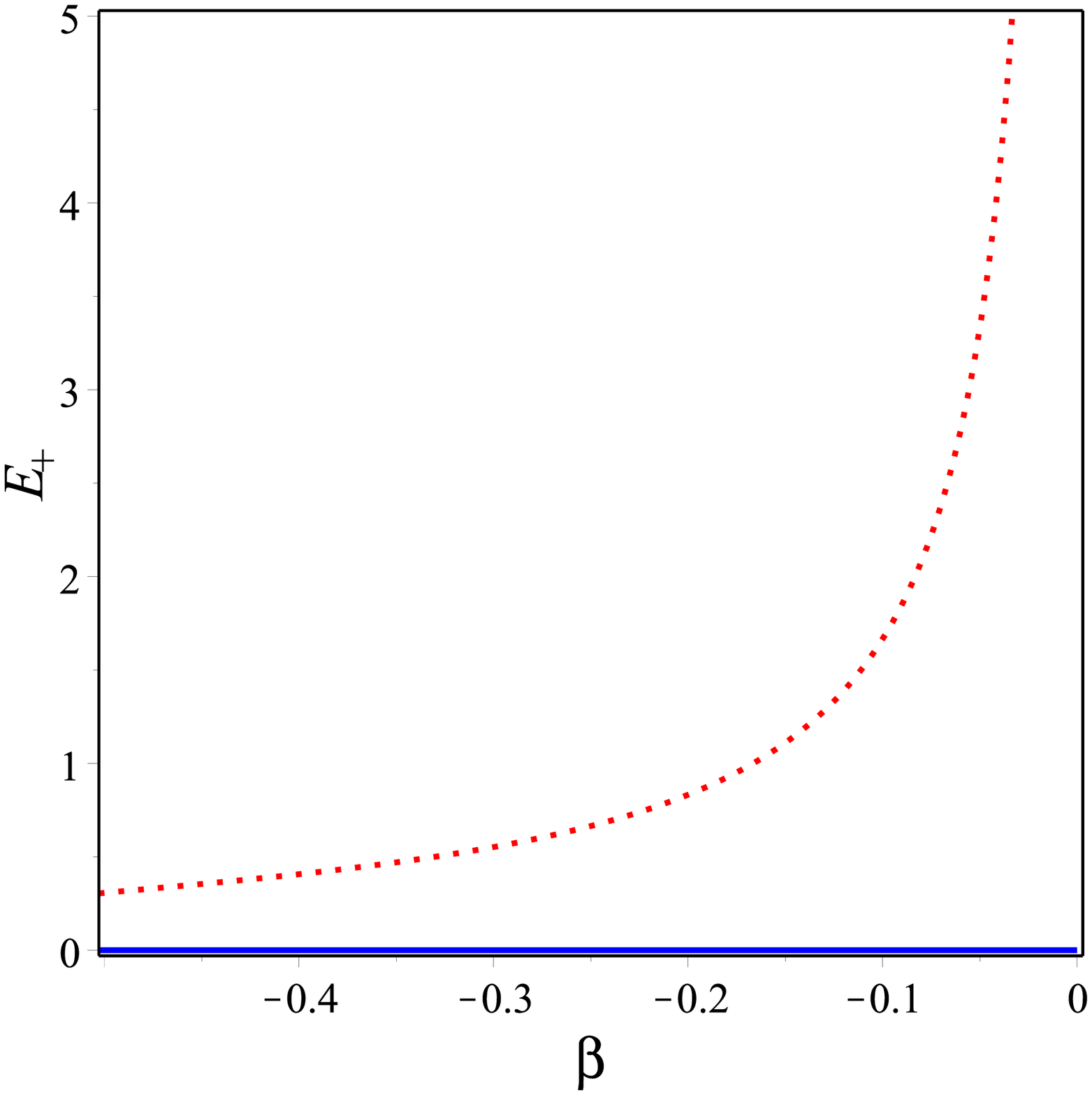}}\hspace{0.2cm}
\subfigure[~The black hole's (\ref{sol1}) quasilocal energy]{\label{fig:4b}\includegraphics[scale=0.4]{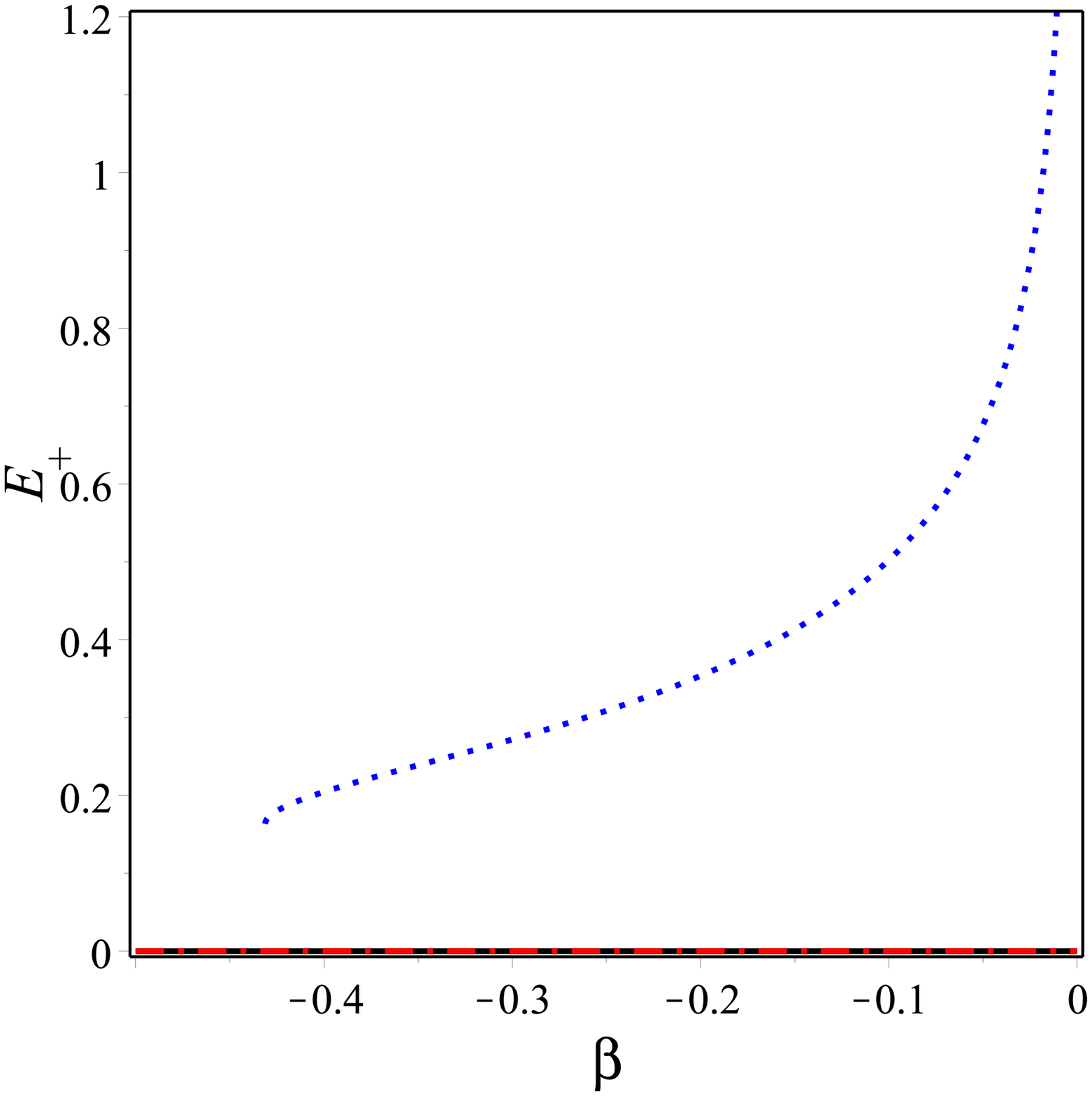}}
\caption{{Plot of the black holes quasilocal energy in the Jordan frame versus $\beta$.}}
\label{Fig:4}
\end{figure}
\begin{figure}
\centering
\subfigure[~The black hole's (\ref{sol}) free energy  ]{\label{fig:5a}\includegraphics[scale=0.4]{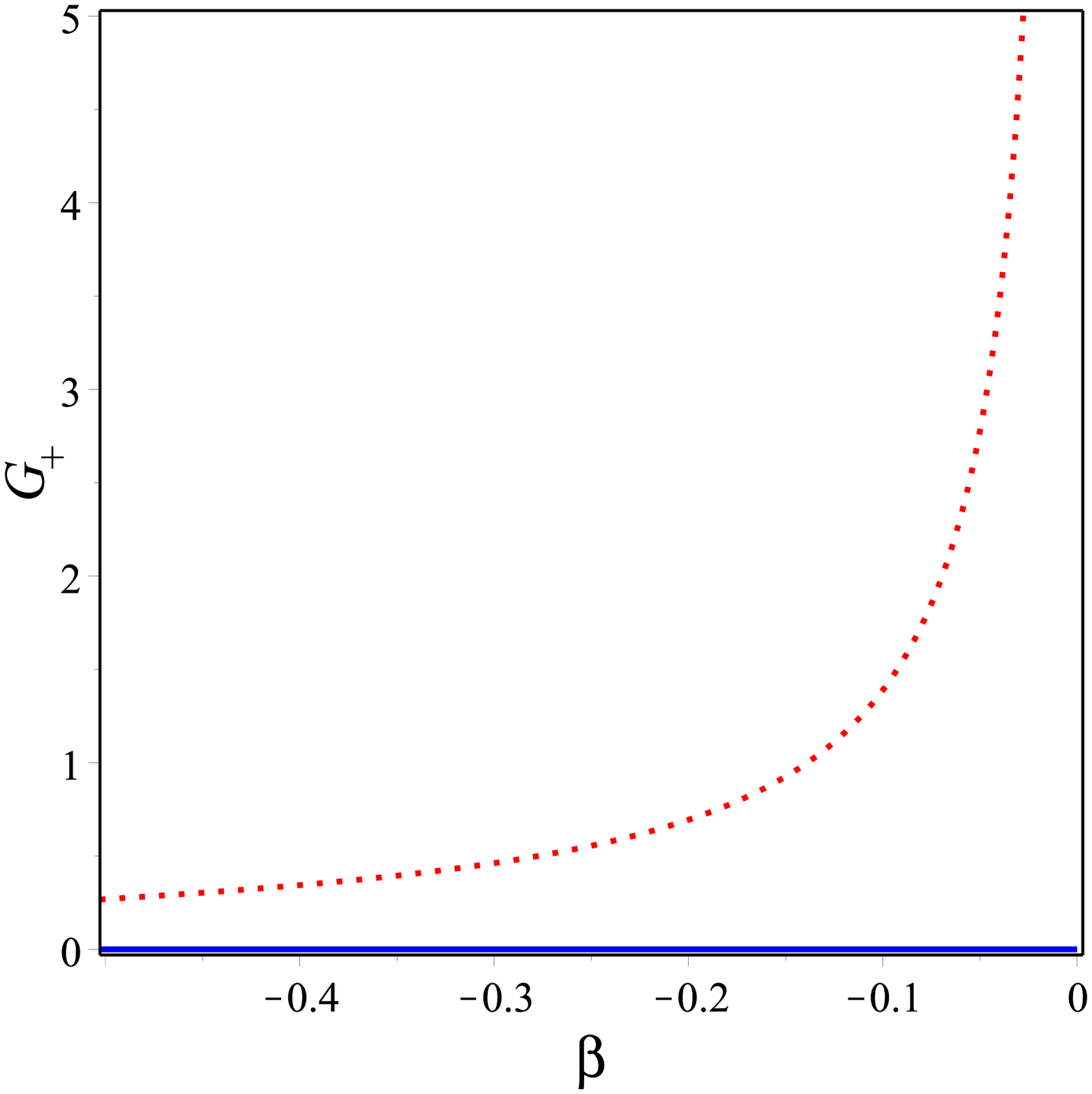}}\hspace{0.2cm}
\subfigure[~The black hole's (\ref{sol1}) free energy ]{\label{fig:5b}\includegraphics[scale=0.4]{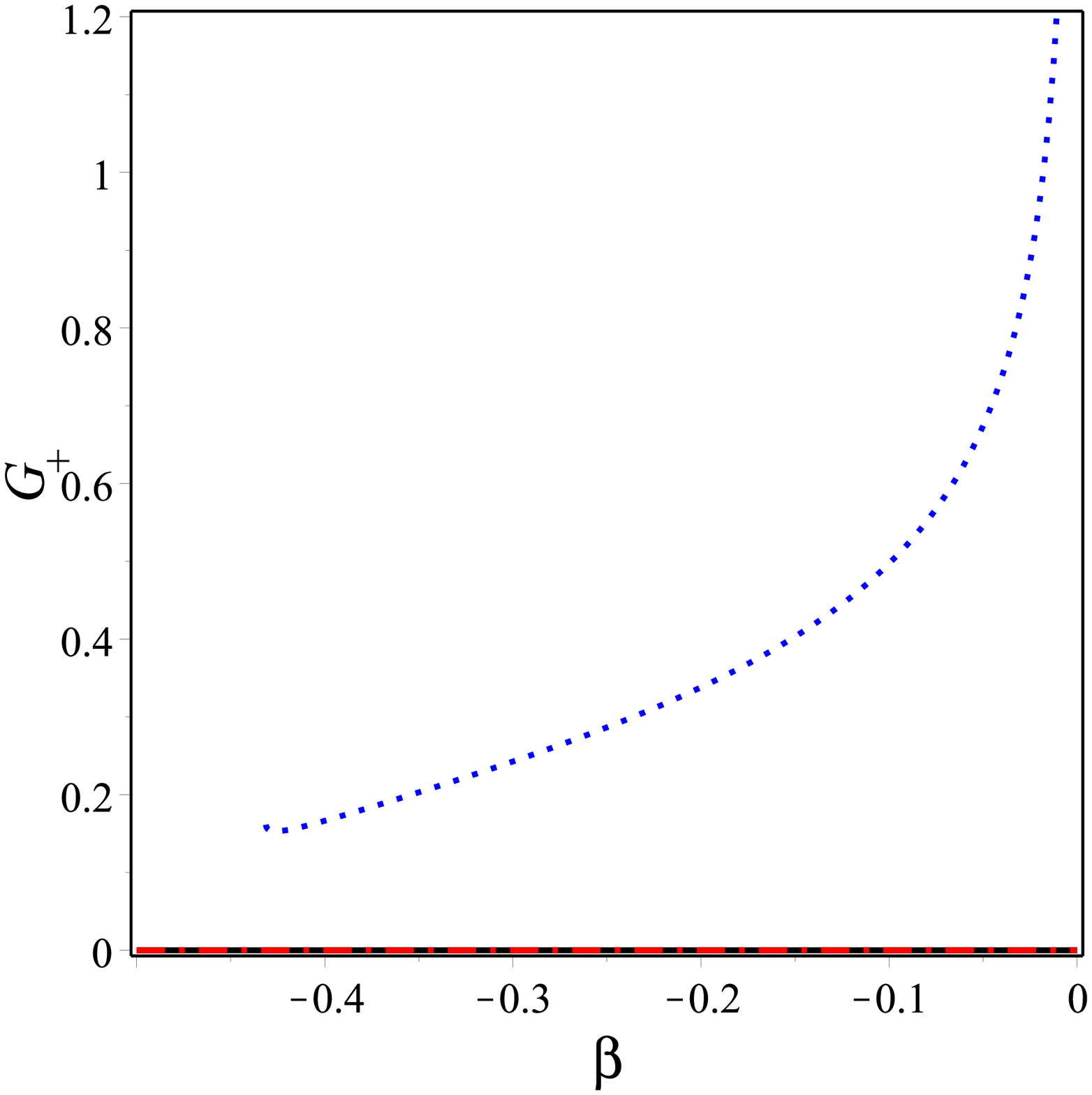}}
\caption{ {Plot of the black hole's  free energy in the Jordan frame versus $\beta$.}}
\label{Fig:5}
\end{figure}
  From the first of Eqs. (\ref{m444}) we see that   $\beta\neq 0$.
The  Gibbs free energy is given by \cite{Zheng:2018fyn,Kim:2012cma}
\begin{equation} \label{enr1}
G(r_+)=E(r_+)-T(r_+)S(r_+)
\end{equation}
where $V$ is the black hole's geometric volume and $P$ is the pressure,
represented by the radial    equation of (\ref{f1}), namely $I_r{}^r$. The quantities
 $E(r_+)$, $T(r_+)$ and $S(r_+)$  are the quasilocal energy, temperature and entropy  at the event horizon, respectively.  From Eqs. (\ref{ent1}),
(\ref{m44e}) and (\ref{m444}) in (\ref{enr1}), we obtain
\begin{eqnarray} \label{m77}
&&{G_+}_{{}_{{}_{{}_{{}_{\tiny Eq. (\ref{sol})}}}}}=\frac{(5+9\beta^2{\cal K}^2)\sqrt{1-18\beta^2{\cal K}^2}-5+9\beta^2{\cal K}^2}{36\beta(1-\sqrt{1-18\beta^2{\cal K}^2})}, \nonumber\\
 &&{G_+}_{{}_{{}_{{}_{{}_{\tiny Eq. (\ref{sol1})}}}}}=\frac{r_+(4+3\beta r_+)}{8}+\frac{(r_++2\beta[2\Lambda \beta r_+{}^4+3\beta {\cal K}^2])(1+\beta r_+)}{12\beta r_+}.
\end{eqnarray}
 In Figs.~\ref{Fig:5}\subref{fig:5a}, \ref{Fig:5}\subref{fig:5b}, the behavior of the black holes' Gibbs energy is represented for particular values of the  parameters of the model.

\subsection{Black hole thermodynamics in Einstein's frame}\label{S666}
In this section we will to repeat the previous calculations but this time in Einstein's frame, i.e. using Eqs. (\ref{conf})  to derive the thermodynamics of the black holes and compare them with the corresponding ones in (\ref{sol}) and (\ref{sol1}).

\begin{figure}
\centering
 \subfigure[~Spherically symmetric space-time]{\label{fig:6a}\includegraphics[scale=0.4]{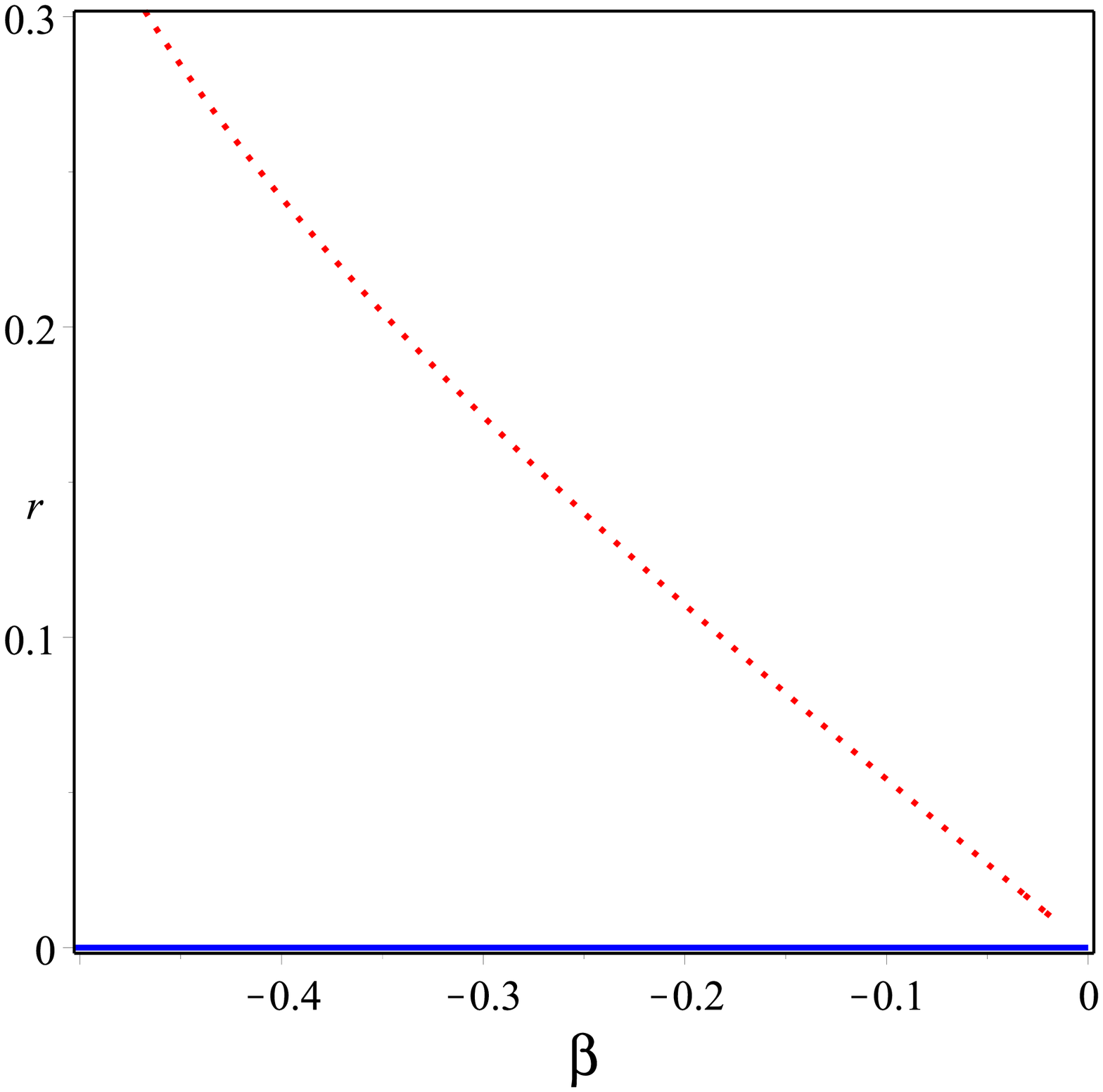}}\hspace{0.2cm}
\subfigure[~Spherically symmetric AdS/dS space-time]{\label{fig:6b}\includegraphics[scale=0.4]{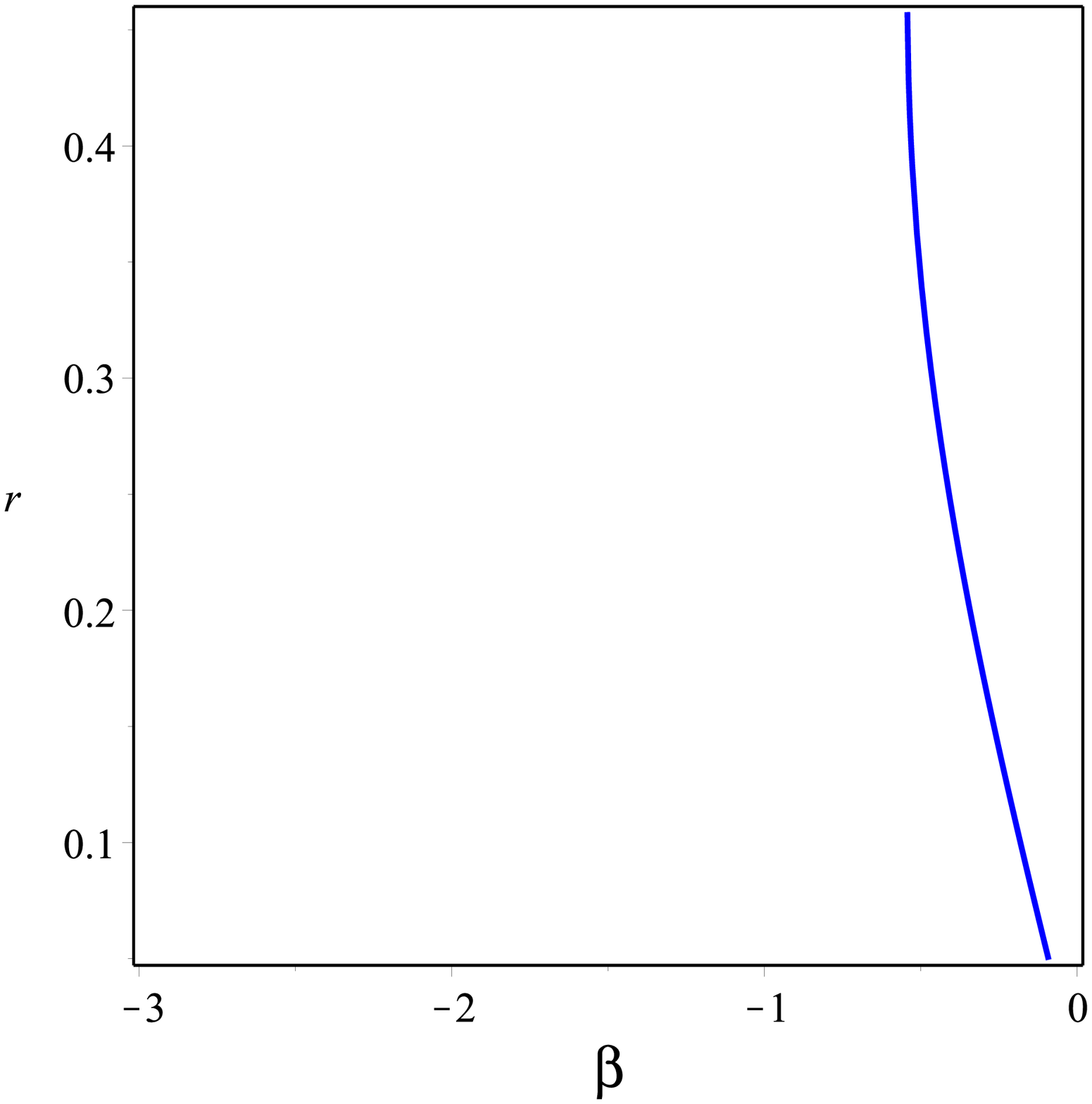}}
\caption{Plot of the radial coordinate  $r$  versus  $\beta$ for the black holes in the Einstein frame.}
\label{Fig:6}
\end{figure}

The constraint $w(r(\bar{r_+})) = 0$ for the flat and AdS/Ad cases gives
\begin{eqnarray} \label{m333}
&&   {\bar r}_+{}_{{}_{{}_{{}_{{}_{\tiny Eq. (\ref{sol})}}}}}=-\frac{\sqrt{6}}{18\beta(2\sqrt{36\beta^2{\cal K}^2+324\beta^4{\cal K}^4-3}-2-36\beta^2{\cal K}^2)^{3/4}}\Big[\sqrt{2\sqrt{36\beta^2{\cal K}^2+324\beta^4{\cal K}^4-3}-2-36\beta^2{\cal K}^2}\Big(1+18\beta^2{\cal K}^2\nonumber\\
& & +\sqrt{36\beta^2{\cal K}^2+324\beta^4{\cal K}^4-3}\Big)-4\Big],
\end{eqnarray}
and for the AdS/dS case one gets an algebraic equation of 8th. order. Eq. (\ref{m333})  shows that  $\beta \neq 0$ cannot be zero, as was already the case in the Jordan frame. Moreover,  Eq. (\ref{m333})  informs  $\beta$ must be negative, such that horizons become positive when there is no charge.  Also Eq. (\ref{m333}) shows that $\beta<\frac{1}{3{\cal K}\sqrt{2}}$ which is consistent with the restriction  put on the dimensional parameter $\beta$ given in the Jordan frame after Eq. (\ref{m33}).  The plot of  $r$ versus  $\beta$  is depicted in Fig. \ref{Fig:6}\subref{fig:6a}. Also, we plot the behavior of $r$ versus $\beta$ for the AdS/dS case in Fig. \ref{Fig:6}\subref{fig:6b}.

\begin{figure}
\centering
\subfigure[~The black hole solution's temperature in Einstein's frame ]{\label{fig:7a}\includegraphics[scale=0.4]{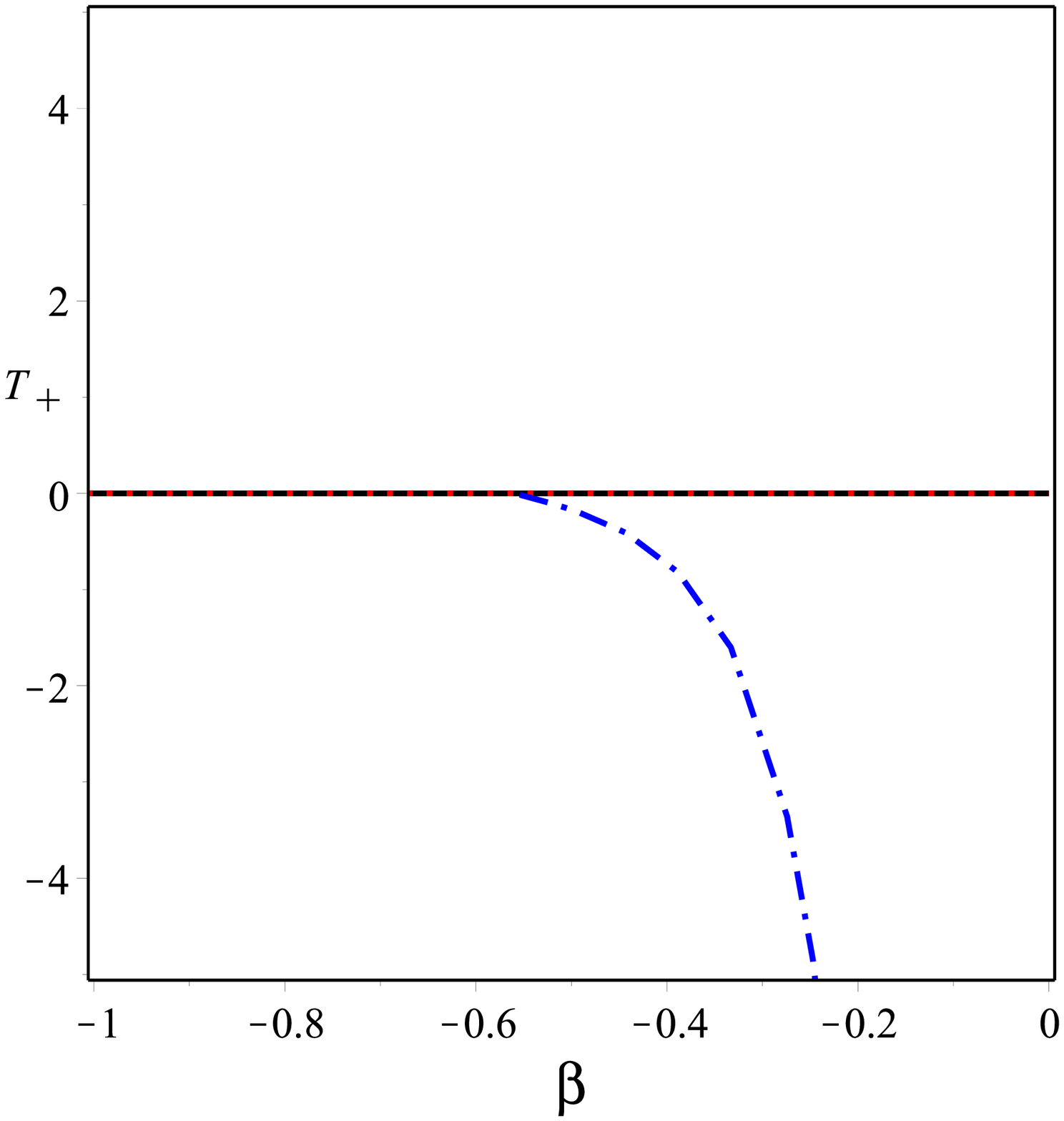}}\hspace{0.2cm}
\subfigure[~The black hole solution's temperature in Einstein's frame]{\label{fig:7b}\includegraphics[scale=0.4]{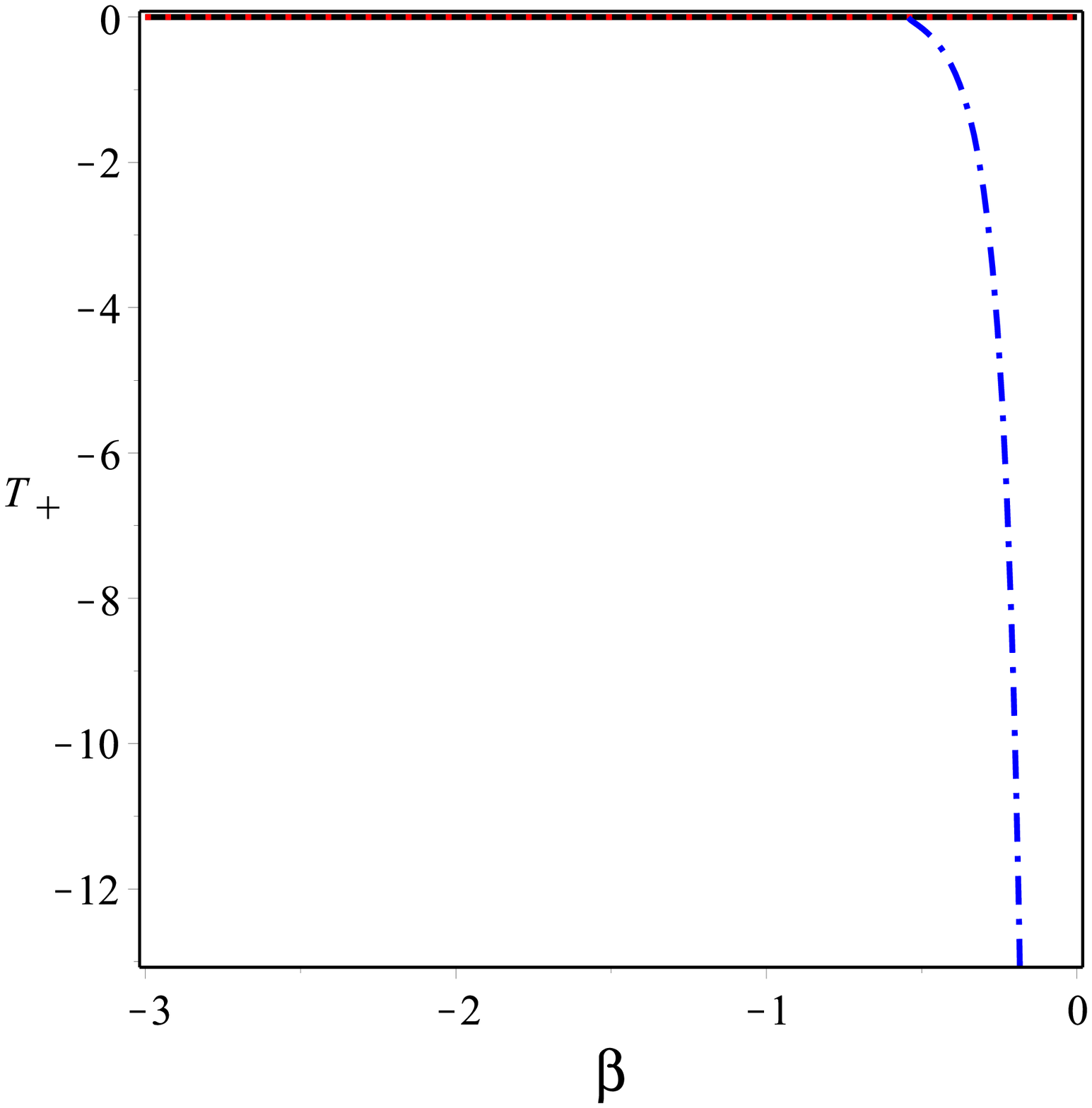}}
\caption{{Plot of the black hole's temperature versus $\beta$ for the black holes in the Einstein frame.}}
\label{Fig:7}
\end{figure}

\begin{figure}
\centering
\subfigure[~The  entropy of solution (\ref{conf})]{\label{fig:8a}\includegraphics[scale=0.4]{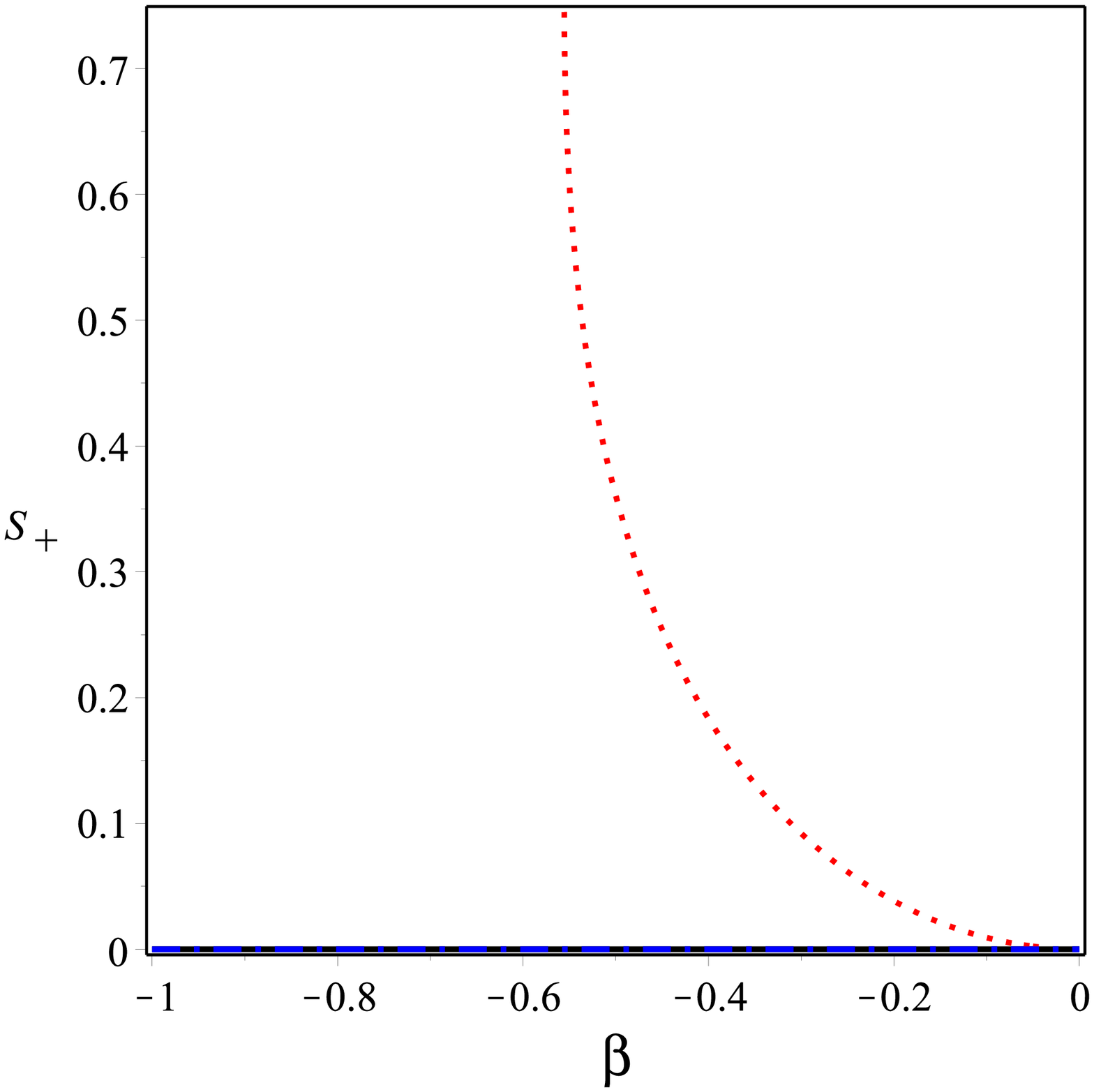}}\hspace{0.2cm}
\subfigure[~The entropy of the AdS/dS  solution (\ref{conf})]{\label{fig:8b}\includegraphics[scale=0.4]{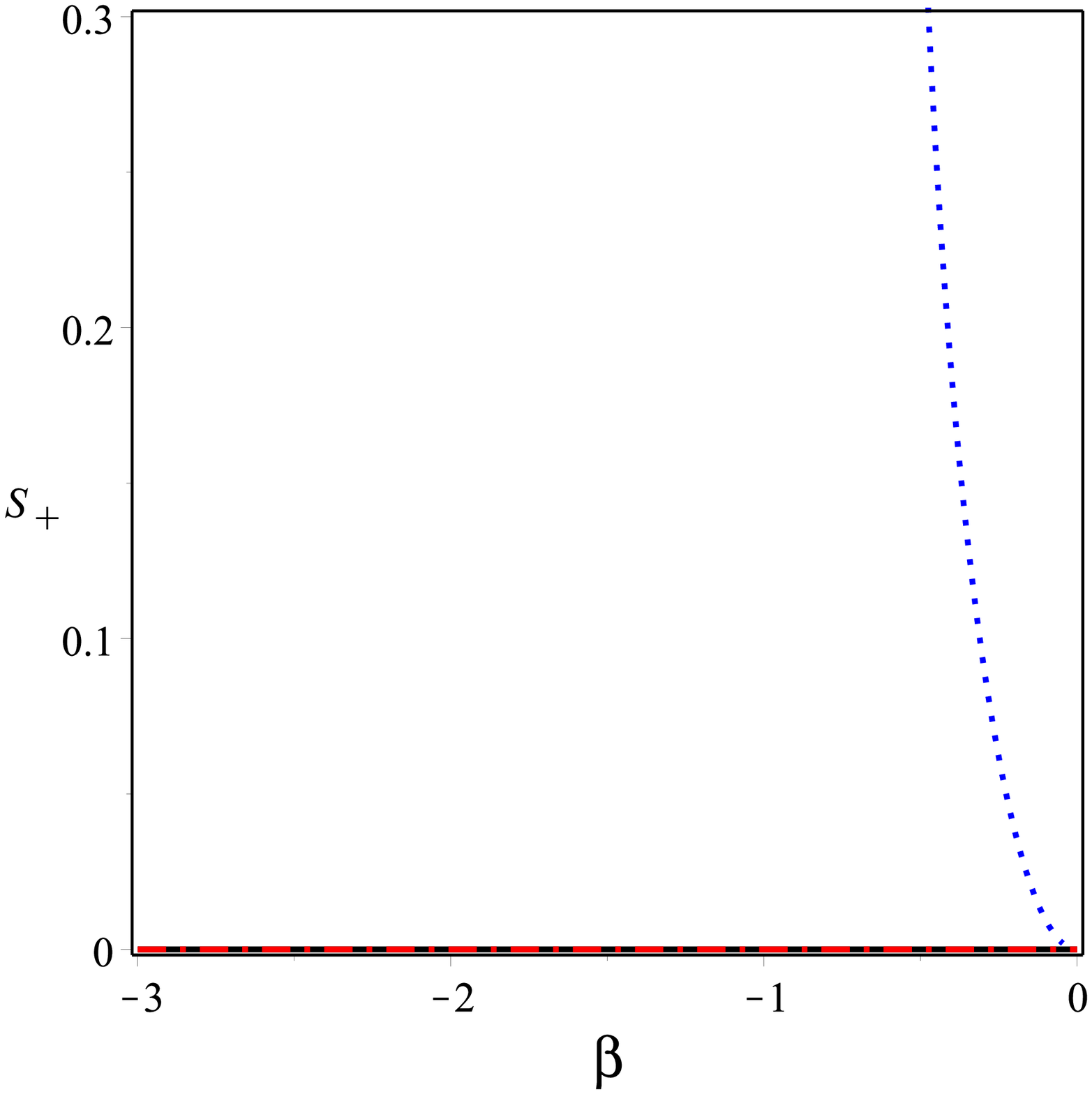}}
\caption{{ Plot of  entropy versus $\beta$  for the black holes in the Einstein frame.}}
\label{Fig:8}
\end{figure}

The Hawking temperature for  each of these black holes  (\ref{conf}) is given by a lengthy expression, but their behaviors can be easily plotted, see Fig.  \ref{Fig:7}.
As is clear from Fig. \ref{Fig:7}, one gets a negative temperature for both black holes in Einstein's frame. If we compare the results of the temperatures in the Jordan and Einstein frames we conclude that the physics of the two frames are not equivalent. This investigation shows in a clear way that  in spite of the equivalence of the two frames from a mathematical viewpoint, and their sharing of many physical properties, the black hole thermodynamics are not equivalent.
The entropy of the black hole (\ref{conf}) in the Einstein frame is defined as
\begin{eqnarray} \label{ente}
{S_+}&=& \pi \bar {r}_+{}^2.
\end{eqnarray}
Using Eq. (\ref{ente}) we compute  the entropy of the solutions (\ref{conf}) as
\begin{eqnarray} \label{ent2}
&&{S_+}= \frac{\pi}{54\beta^2(2\sqrt{36\beta^2{\cal K}^2+324\beta^4{\cal K}^4-3}-2-36\beta^2{\cal K}^2)^{3/2}}\Big[-4+\sqrt{2\sqrt{36\beta^2{\cal K}^2+324\beta^4{\cal K}^4-3}-2-36\beta^2{\cal K}^2}\Big(1+18\beta^2{\cal K}^2\nonumber\\
& & +\sqrt{36\beta^2{\cal K}^2+324\beta^4{\cal K}^4-3}\Big)\Big]^2, \nonumber\\
&&{S_+}_{{}_{{}_{{}_{{}_{AdS/dS}}}}}=\pi  \bar {r}_+{}^2.
\end{eqnarray}
Eqs. (\ref{ent2}) are plotted in Fig. \ref{Fig:8}, showing that we have a positive entropy.  We note that  $S$ is proportional to $A$, because of the fact that we are in Einstein's frame.

\begin{figure}
\centering
\subfigure[~Black hole solution's Quasilocal energy  ]{\label{fig:9a}\includegraphics[scale=0.4]{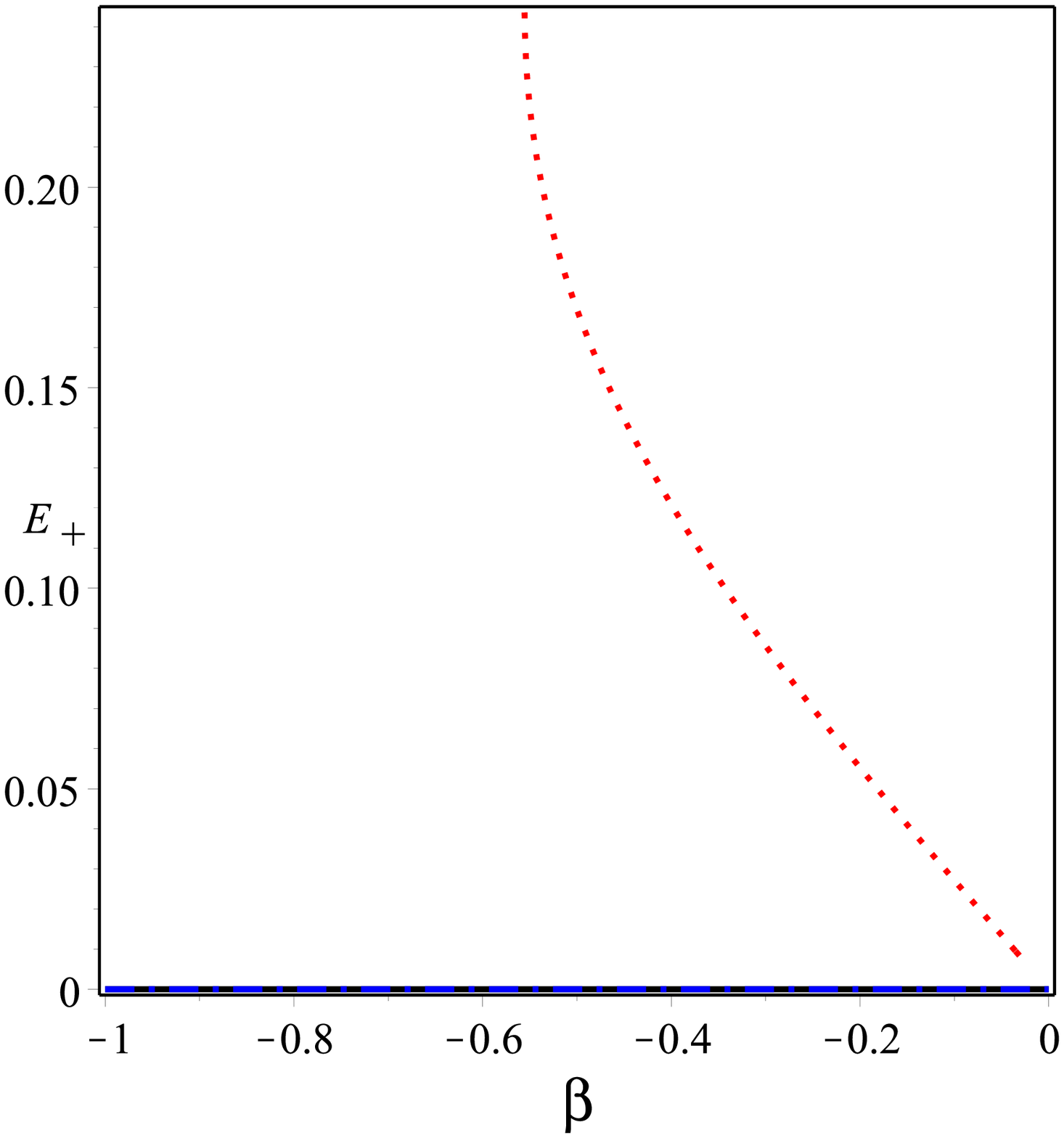}}\hspace{0.2cm}
\subfigure[~Black hole solution's Quasilocal energy ]{\label{fig:9b}\includegraphics[scale=0.4]{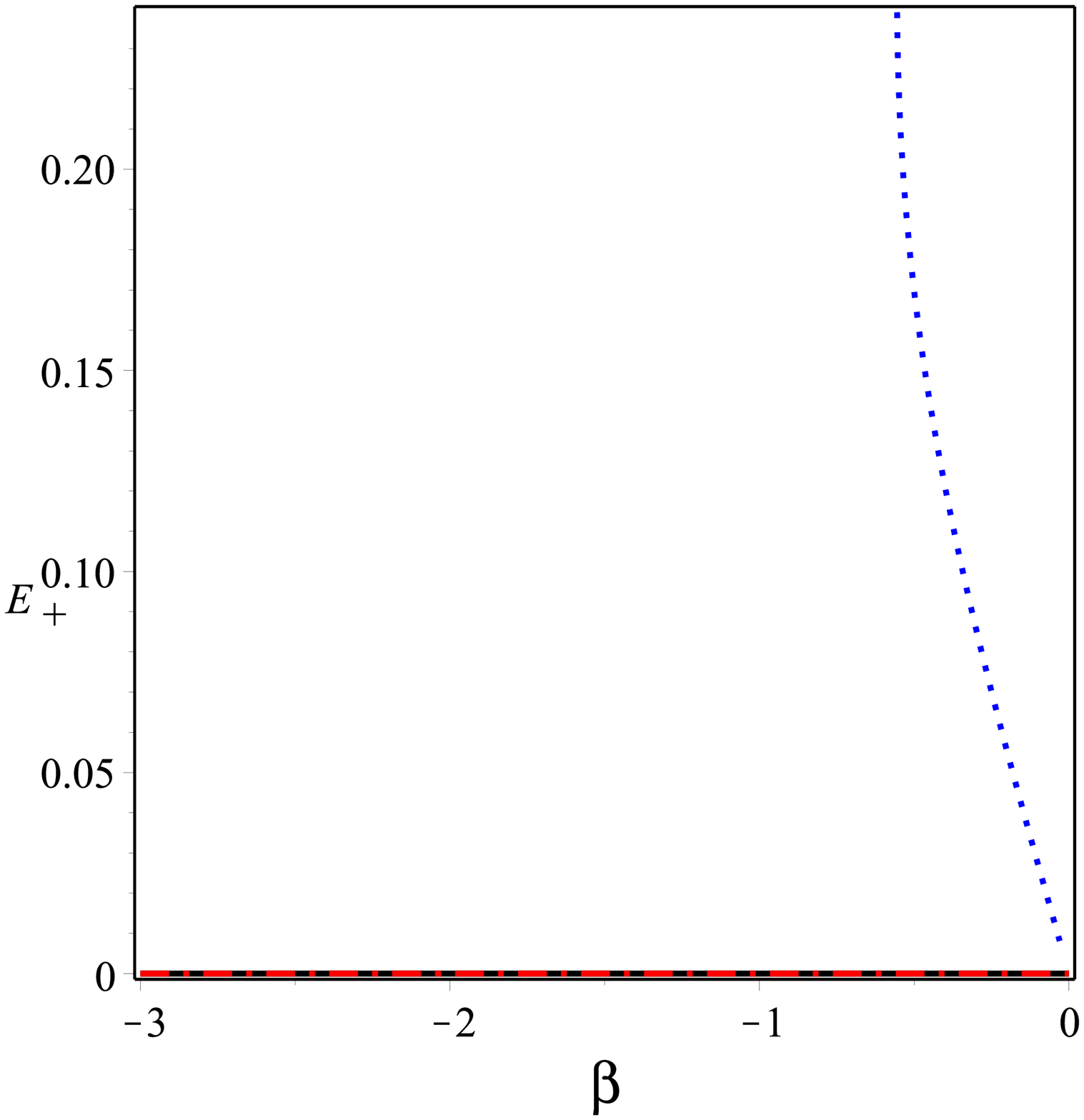}}
\caption{{Plot of the quasilocal energies versus $\beta$ for the black holes in Einstein frame.}}
\label{Fig:9}
\end{figure}

Using Eq. (\ref{en}), the quasi-local energies give
\begin{eqnarray} \label{m44}
{E_+}_{{}_{{}_{{}_{{}_{\tiny Eq. (\ref{sol})}}}}}&=&-\frac{1+9\beta^2 {\cal K}^2-\sqrt{1-18\beta^2 {\cal K}^2}}{12 \beta},\nonumber\\
{E_+}_{{}_{{}_{{}_{{}_{\tiny Eq. (\ref{sol1})}}}}}&=&\frac{\bar {r}_+}{8}\left(4+3\beta \bar {r}_+\right).
\end{eqnarray}
The behavior of the quasi local energy is plotted in Fig. \ref{Fig:9}, which shows that we obtain a positive quasi local energy.
\begin{figure}
\centering
\subfigure[~Black hole solution's free energy  ]{\label{fig:10a}\includegraphics[scale=0.4]{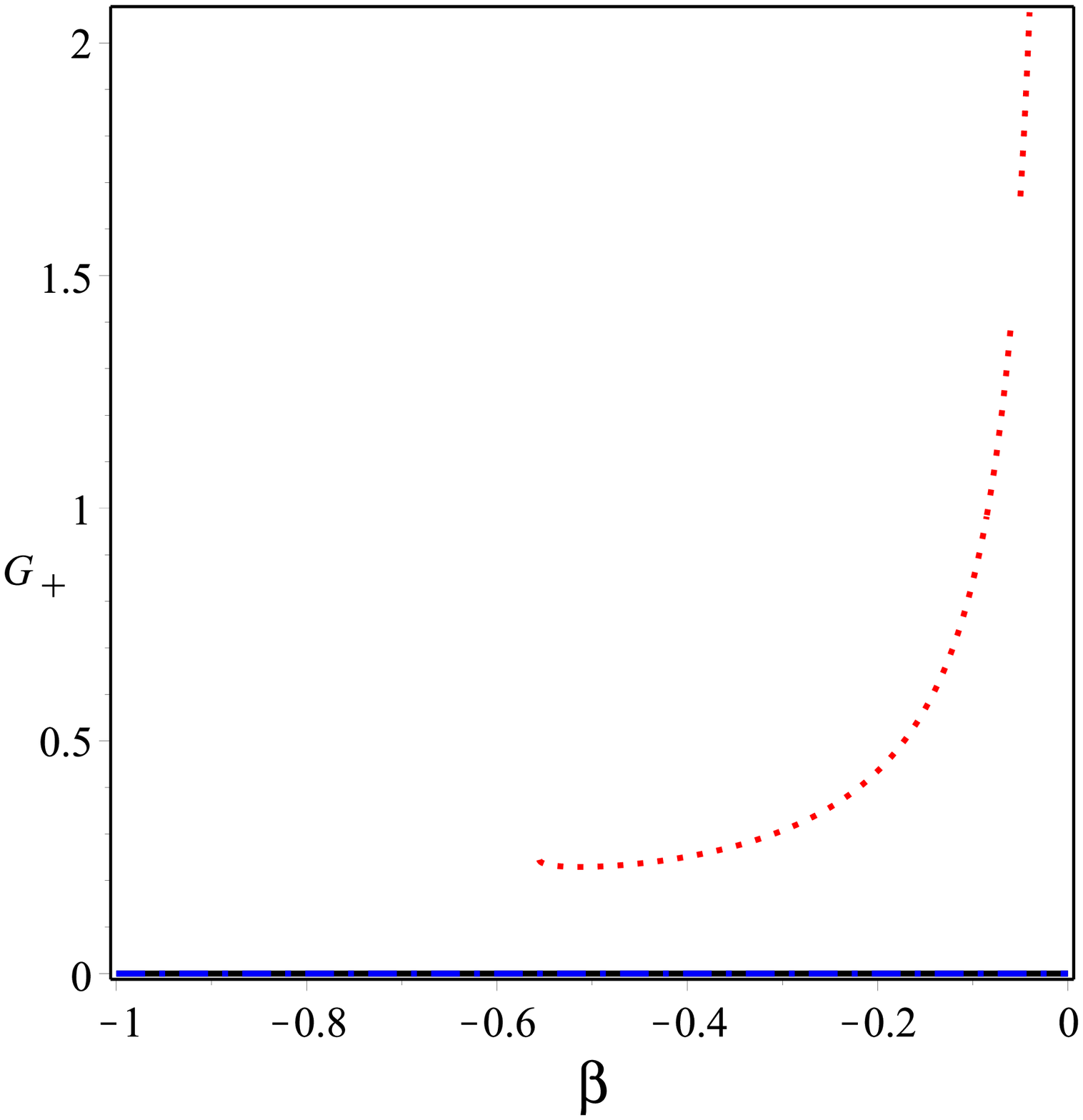}}\hspace{0.2cm}
\subfigure[~Black hole solution's free energy ]{\label{fig:10b}\includegraphics[scale=0.4]{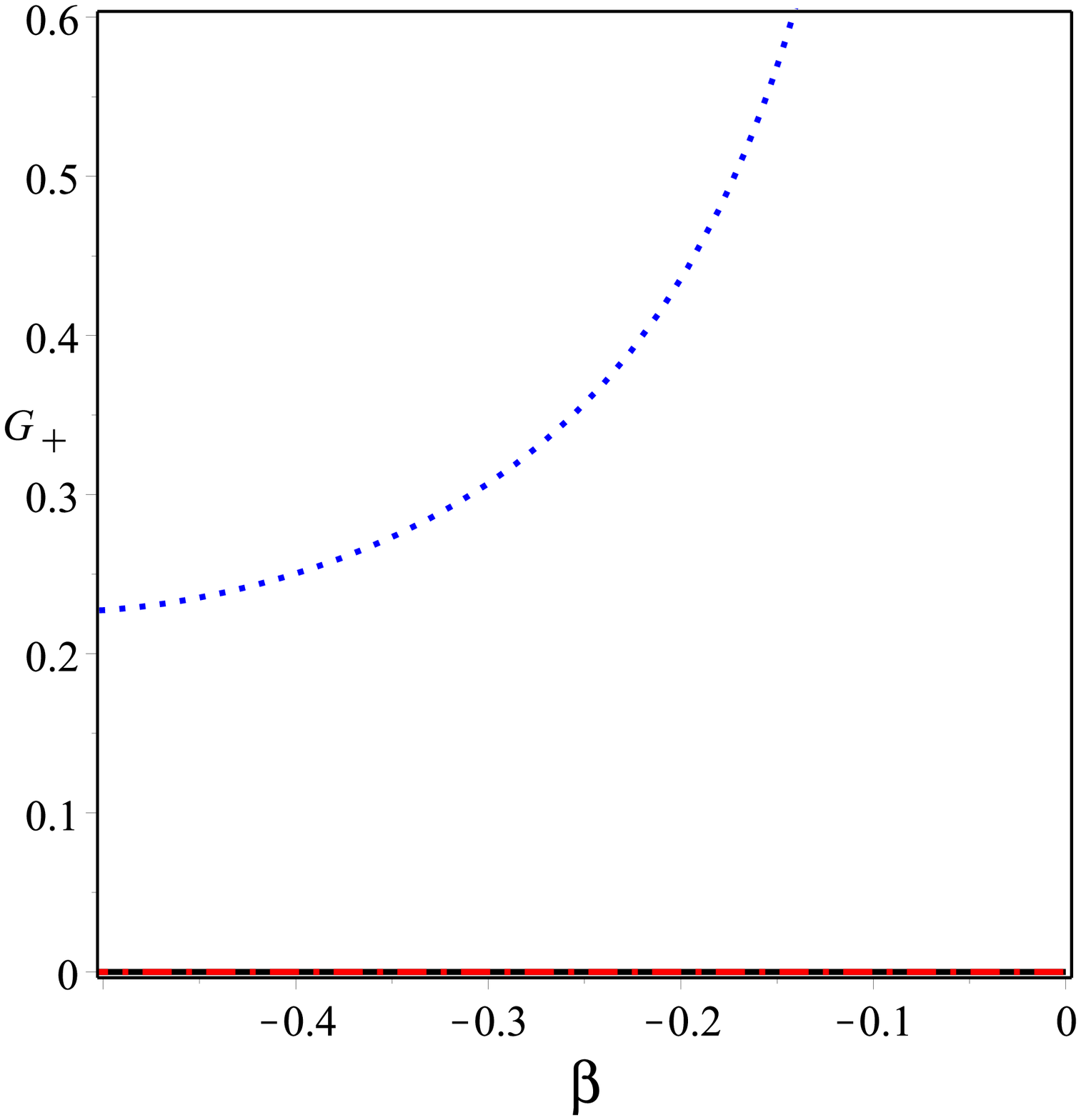}}
\caption{ {Plot of  the free energy versus $\beta$ for solutions (\ref{conf}).}}
\label{Fig:10}
\end{figure}
  Equation  (\ref{m44})  shows that  $\beta\neq 0$.

The free energy  is given by
\begin{equation} \label{enr}
G(r_+)=E({\bar r}_+)-T({\bar r}_+)S(r{\bar r}_+).
\end{equation}
Using Eqs.  (\ref{m44}) and (\ref{ent2}) in (\ref{enr}), we get
\begin{eqnarray} \label{m77}
&&{G_+}_{{}_{{}_{{}_{{}_{\tiny Eq. (\ref{sol})}}}}}=\frac{(5+9\beta^2{\cal K}^2)\sqrt{1-18\beta^2{\cal K}^2}-5+9\beta^2{\cal K}^2}{36\beta(1-\sqrt{1-18\beta^2{\cal K}^2})}, \nonumber\\
 &&{G_+}_{{}_{{}_{{}_{{}_{\tiny Eq. (\ref{sol1})}}}}}=\frac{{\bar r}_+(4+3\beta {\bar r}_+)}{8}+\frac{({\bar r}_++2\beta[2\Lambda \alpha {\bar r}_+{}^4+3\beta {\cal K}^2])(1+\beta {\bar r}_+)}{12\beta {\bar r}_+}.
\end{eqnarray}
The behavior of the functions in (\ref{m77})  is depicted in Figs.~\ref{Fig:10}\subref{fig:10a}, \ref{Fig:10}\subref{fig:10b} for particular values of the  parameters of the model.
\section{Stability analysis of the black holes in the Jordan and Einstein frames}\label{S626}
To study the stability of the black holes we rewrite rewrite Eq. (\ref{a2}) as
\begin{equation}
{\cal S}=\frac{1}{2\kappa}\int d^{4}x\sqrt{-g}\,[\psi\, {\cal R}-V(\psi)]\, , \label{action2}
\end{equation}
where we  neglect  $\Lambda$.
Here the scalar field  $\psi$ is coupled to the Ricci scalar and $V(\psi)$ is the potential field  \cite{Capozziello:2011et}.
Discussion of the stability of the solutions derived in the previous sections is done through the study of perturbations of spherically symmetric  vacuum background spacetime, endowed with the   metric
\begin{equation} {\mathrm ds^2}=g_{\mu\nu}^{BG}dx^{\mu}dx^{\nu}=-w(r)\, dt^{2}+\frac{dr^{2}}{w_1(r)}+r^{2} ( d\theta^{2} +\sin^2\theta\, d\phi^{2}),
\end{equation}
 $g_{\mu\nu}^{BG}$ being the background metric. We investigate the stability of the black holes obtained in the Jordan and Einstein frames proceeds by using linear perturbations. We have
\begin{equation}
  V =- {\frac{4w_1\, \psi'}{r}}-{\frac{2\psi\, w'w_1}{w r}}-\frac{\psi' w'w_1}{w}+{\frac{2\psi}{{r}^{2}}}-{\frac{2w_1\, \psi}{{r}^{2}}}\, , \qquad \psi'' =-\frac{w'_1\psi'}{2w_1}-\frac{\psi w'_1}{rw_1}+\frac{w'\psi'}{2w}+\frac{\psi w'}{rw}\,, \qquad   {\cal R} =\frac{dV}{d\psi}\,,\label{eq:Ueq}\\
\end{equation} as  background
equation of motion in which  $'$ stands for differentiation w.r.t $r$.

\subsection{Brief review of the Regge-Wheeler-Zerilli prescription}

We shall now give an outline  prescription of the Regge, Wheeler \cite{PhysRev.108.1063}, and Zerilli \cite{PhysRevLett.24.737}, which can also
be used in modified $f({\cal R})$ gravity \cite{Nashed:2019tuk}. We start from the slightly perturbed  metric corresponding to a static spherically symmetric  space-time, $g_{\mu\nu}=g_{\mu\nu}^{BG}+h_{\mu\nu}$,  where $h_{\mu\nu}$ stands for
 an infinitesimal quantity. A scalar  field $\Psi(t,r,\theta,\phi)$ can be decomposed as
 \begin{equation}
\Psi(t,r,\theta,\phi)=\sum_{\ell,m}\Phi_{\ell m}(t,r)Y_{\ell m}(\theta,\varphi),\label{scalar-decomposition}
\end{equation}
where $Y_{\ell m}(\theta,\phi)$ are spherical harmonics.
In the same way, one can decompose any vector $V_{a}$  into a  divergent and a non-divergent parts, i.e.,
 \begin{equation} \label{v77}
V_{a}(t,r,\theta,\phi)=\nabla_{a}\Psi_{1}+E_{a}^b\nabla_b\Psi_{2},
\end{equation}
with $\Psi_{1}$ and $\Psi_{2}$ being two scalars and $E_{ab}\equiv\sqrt{\det\gamma}~\epsilon_{ab}$,
where $\gamma_{ab}$ is a 2-dimension metric
and $\epsilon_{ab}$  the usual  anti-symmetric tensor, where
$\epsilon_{\theta \varphi}=1$. The symbol $\nabla_{a}$ stands for the covariant differentiation w.r.t.  $\gamma_{ab}$.
Eq. (\ref{v77}) shows for the two-component vector   $V_{a}$,  it is possible
 to specify it  by the two scalar quantities $\Psi_{1}$ and $\Psi_{2}$. Therefore, one can use Eq. (\ref{scalar-decomposition}) with $\Psi_{1}$
and $\Psi_{2}$ in order to express the vector quantity $V_a$ in spherical
harmonics.

On the other hand, any symmetric tensor $T_{ab}$  can be rewritten as
 \begin{equation}
T_{ab}(t,r,\theta,\phi)=\nabla_{a}\nabla_{b}\Psi_{1}+\gamma_{ab}\Psi_{2}+\frac{1}{2}\left(E_{a}{}^{c}\nabla_{c}\nabla_{b}\Psi_{3}+
E_{b}{}^{c}\nabla_{c}\nabla_{a}\Psi_{3}\right),
\end{equation}
where $\Psi_{1},~\Psi_{2}$ and $\Psi_{3}$ are three scalar quantities since $T_{ab}$ has three
independent components. Therefore, one can use the scalar decomposition
(\ref{scalar-decomposition}) with $\Psi_{1},~\Psi_{2}$ and $\Psi_{3}$, to determine the tensor $T_{ab}$. The variables corresponding to $E_{ab}$ are the
 odd-type ones while the rest represent the ones of even-type.  The quantities $h_{\mu\nu}$ in the linearized  form make
odd and even perturbations to fully decouple which makes the above procedure useful to use. In the following subsection we are going to study the perturbations of odd-type.

\subsection{Perturbative form of $f({\cal R})$ theory}

\centerline{The odd-types perturbations}
From the Regge-Wheeler method, the  metric perturbations of odd-type take the form  \cite{PhysRev.108.1063,PhysRevLett.24.737}
 \begin{eqnarray}
 &  & h_{tt}=0,~~~h_{tr}=0,~~~h_{rr}=0,\\
 &  & h_{ta}=\sum_{\ell, m}h_{0,\ell m}(t,r)E_{ab}\partial^{b}Y_{\ell m}(\theta,\varphi),\\
 &  & h_{ra}=\sum_{\ell, m}h_{1,\ell m}(t,r)E_{ab}\partial^{b}Y_{\ell m}(\theta,\varphi),\\
 &  & h_{ab}=\frac{1}{2}\sum_{\ell, m}h_{2,\ell m}(t,r)\left[E_{a}^{~c}\nabla_{c}\nabla_{b}Y_{\ell m}(\theta,\varphi)+E_{b}^{~c}\nabla_{c}\nabla_{a}Y_{\ell m}(\theta,\varphi)\right]. \label{pert}
\end{eqnarray}
Using the gauge transformation $x^{\mu}\to x^{\mu}+\xi^{\mu}$, where the components $\xi^{\mu}$ are infinitesimal, one can prove that some components of the metric
perturbations are not physical and can be put equal to zero. Now we consider, for the odd perturbations, the following transformation
 \begin{equation}
\xi_{t}=\xi_{r}=0,\qquad \qquad \xi_{a}=\sum_{\ell m}\Lambda_{\ell m}(t,r)E_{a}^{~b}\nabla_{b}Y_{\ell m},
\end{equation}
where $\Lambda_{\ell m}$  can always be set to vanish. Through this method one can show that  $\Lambda_{\ell m}$ is
fully fixed which means that  they are free from any gauge degrees of freedom. Using Eq. (\ref{pert}) in
Eq.  (\ref{action2}) and integrating by parts Eq. (\ref{action2})  becomes
\begin{eqnarray}
&&S_{{ odd}}=\frac{1}{2\kappa} \sum_{\ell,m}\int dt\, dr\,{\cal L}_{{ odd}}=\frac{1}{4\kappa} \sum_{\ell,m}\int dt\, dr\,j^{2}\Bigg[\frac{\phi \sqrt{w_1}}{\sqrt{w}}\Big\{{\left({\dot{h}_{1}}-h_{0}'\right)}^{2}+\frac{8h_{0}{\dot{h}_{1}}}{r}\Big\}+ \frac{h_{0}^{2}}{r^{2}}\left\{4r\Bigg[\frac{\phi \sqrt{w_1}}{\sqrt{w}}\Bigg]'+4\frac{\phi \sqrt{w_1}}{\sqrt{w}} +\frac{(j^2-2)\phi}{\sqrt{ww_1}} \right\}\nonumber\\
&&-\frac{\,(j^{2}-2)\,\sqrt{ww_1}\, \phi\, h_{1}^{2}}{r^{2}}\Bigg],\label{odd-action}
\end{eqnarray}
where we have dropped the suffix $\ell$ for the fields, and $j^{2}=\ell\,(\ell+1)$. Variation of (\ref{odd-action}) w.r.t. $h_{0}$ yields
 \begin{equation}
\Bigg[\phi\sqrt{\frac{w_1}{w}}(h_{0}'-\dot{h}_{1})\Bigg]'= \frac{ h_{0}}{r^{2}}\left\{4r\Bigg[\frac{\phi \sqrt{w_1}}{\sqrt{w}}\Bigg]'+4\frac{\phi \sqrt{w_1}}{\sqrt{w}} +\frac{(j^2-2)\phi}{\sqrt{ww_1}} \right\}+\frac{4\phi\sqrt{\frac{w_1}{w}}\,\dot{h}_{1}}{r}\,,\label{cons}
\end{equation}
that cannot be solved for $h_{0}$. Therefore, we are going to rewrite the action (\ref{odd-action}) as
\begin{equation}
{L}_{{ odd}}=\frac{j^{2}\, \phi \sqrt{w_1}}{2\sqrt{w}}{\left({\dot{h}_{1}}-h_{0}'+\frac{2\,{h_0}}{r}\right)}^{2}-\frac{j^{2}\Bigg(\frac{\phi \sqrt{w_1}}{\sqrt{w}}+r\Bigg\{\frac{\phi \sqrt{w_1}}{\sqrt{w}}\Bigg\}'\Bigg)\,h_{0}{}^2}{r^2}+\frac{j^{2}\,h_{0}^{2}}{r^{2}}\left\{4r\Bigg[\frac{\phi \sqrt{w_1}}{\sqrt{w}}\Bigg]'+4\frac{\phi \sqrt{w_1}}{\sqrt{w}} +\frac{(j^2-2)\phi}{\sqrt{ww_1}} \right\}-\frac{j^2\,(j^{2}-2)\,\sqrt{ww_1}\, \phi\, h_{1}^{2}}{r^{2}}.\label{eq:Lodd2}
\end{equation}
In Eq. (\ref{eq:Lodd2})  all expressions involving  $\dot{h}_{1}$ are collected in first term. Using the Lagrange multiplier,  Eq.~(\ref{eq:Lodd2}) becomes
\begin{eqnarray}
&&{L}_{{  odd}}=\frac{j^{2}\, \psi \sqrt{w_1}}{2\sqrt{w}}\left[2\, Q\left(\dot{h}_{{1}}-h'_{{0}}+{\frac{2\,h_{{0}}}{r}}\right)-Q^{2}\right]-\frac{j^{2}\Bigg(\frac{\psi \sqrt{w_1}}{\sqrt{w}}+r\Bigg\{\frac{\psi \sqrt{w_1}}{\sqrt{w}}\Bigg\}'\Bigg)\,h_{0}{}^2}{r^2}+\frac{j^{2}\,h_{0}^{2}}{r^{2}}\left\{4r\Bigg[\frac{\psi \sqrt{w_1}}{\sqrt{w}}\Bigg]'+4\frac{\psi \sqrt{w_1}}{\sqrt{w}} +\frac{j^2(j^2-2)\phi}{\sqrt{ww_1}} \right\}\nonumber\\
&&-\frac{j^2\,(j^{2}-2)\,\sqrt{ww_1}\, \phi\, h_{1}^{2}}{r^{2}}.\label{eq:Lodd3}
\end{eqnarray}
Eq.~(\ref{eq:Lodd3}) shows that the fields, $h_{0}$ and $h_{1}$, take the form
\begin{eqnarray}
h_{1} & = & -\frac{r^2\,\dot{Q}}{(j^{2}-2)w}\,,\label{eq:oddh1}\\
h_{0} & = & \frac{r\Bigg[\Bigg(\frac{2\phi \sqrt{w_1}}{\sqrt{w}}+r\Bigg\{\frac{\psi \sqrt{w_1}}{\sqrt{w}}\Bigg\}'\Bigg)\, Q+\frac{r\, \psi \sqrt{w_1}}{2\sqrt{w}} Q'\Bigg]}{2\Bigg(\frac{\psi \sqrt{w_1}}{\sqrt{w}}+r\Bigg\{\frac{\psi \sqrt{w_1}}{\sqrt{w}}\Bigg\}'\Bigg)-\Bigg(\frac{\psi \sqrt{w_1}}{\sqrt{w}}+r\Bigg\{\frac{\psi \sqrt{w_1}}{\sqrt{w}}\Bigg\}'\Bigg)}\:.\label{eq:oddh0}
\end{eqnarray}
Eqs.  (\ref{eq:oddh1}) and (\ref{eq:oddh0}) relate  $h_{0}$ and $h_{1}$ to
the auxiliary field $Q$. When we know  $Q$ the physical modes $h_{0}$ and $h_{1}$
become known. Substituting Eqs. (\ref{eq:oddh1}) and (\ref{eq:oddh0}) into Eq. (\ref{eq:Lodd3}) and after some manipulation we get
\begin{equation}
{L}_{{odd}}=\frac{s_1{}^2}{s_2}\,\dot{Q}^{2}-\frac{s_1{}^2\,r^{2}}{s_3r^2-2rs'_1-2s_1}\,Q'^{2}-\nu_1{}^{2}\, Q^{2}\,,\label{eq:LoddF}
\end{equation}
 where
\begin{eqnarray}
&& s_1=\frac{j^2\psi \sqrt{w_1}}{2\sqrt{w}},\qquad s_2=\frac{j^2\psi(j^2-2)\sqrt{ww_1}}{2r^2}, \qquad s_3=\frac{j^2}{r^2}\Bigg(\frac{\psi \sqrt{w_1}}{\sqrt{w}}+r\Bigg\{\frac{\psi \sqrt{w_1}}{\sqrt{w}}\Bigg\}'+\frac{(j^2-2)\psi}{2\sqrt{ww_1}}\Bigg),\nonumber\\
&&\nu_1{}^{2}=\frac{s_1r^2\Big[r^2s'_1s'_3-r^2s''_1s_3+2s_1s_3+4s'_1{}^2+r^2s_3{}^2-2s_1s''_1+2rs_1s'_3-4rs'_1s_3\Big]}{(2s_1+2rs'_1-r^2s_3)^2}\,.
\end{eqnarray}
Using  Eq.~(\ref{eq:LoddF}), one obtains
\[
s_2\geq2\,,\qquad{  \mbox{that\, leads\, to} \, \,\qquad}j^2\geq2\,,
\]
which is  the no ghost conditions.
One can derive the radial dispersion relation as
\[
\omega^{2}=ww_1\, k^{2}\,.
\]
 In the above equation we have assumed that the solutions proportional to $e^{i(\omega t-kr)}$  where
$k$ and $\omega$ are large. For the radial speed we get
\[
c_{{ odd}}^{2}=\left(\frac{dr_{*}}{d\tau}\right)^2=1\,,
\]
 where  the radial tortoise coordinate ($dr_{*}^{2}=dr^{2}/w_1$)
and the proper time ($d\tau^{2}=w_1\, dt^{2}$) have been employed.

\section{Black hole stability analysis using geodesic deviations in Jordan's frame}\label{S9}
 The paths  of a test particle in the gravitational field are described by
 \begin{equation}\label{ge}
 {d^2 x^\sigma \over d\tau^2}+ \left\{^\sigma_{ \mu \nu} \right \}
 {d x^\mu \over d\tau}{d x^\nu \over d\tau}=0,
 \end{equation}
which is known as the geodesic equations.  In Eq. (\ref{ge})  $\tau$ represents  the affine  connection parameter. The
  geodesic  deviation  has the form \cite{1992ier..book.....D,Nashed:2003ee}
  \begin{equation} \label{ged}
 {d^2 \xi^\sigma \over d\tau^2}+ 2\left\{^\sigma_{ \mu \nu} \right \}
 {d x^\mu \over d\tau}{d \xi^\nu \over d\tau}+
 \left\{^\sigma_{ \mu \nu} \right \}_{,\ \rho}
 {d x^\mu \over d\tau}{d x^\nu \over d\tau}\xi^\rho=0,
 \end{equation}
where $\xi^\rho$ is the 4-vector deviation. Introducing (\ref{ge}) and (\ref{ged})  into (\ref{met}), one can get
 \begin{equation}
{d^2 t \over d\tau^2}=0, \qquad {1 \over 2} w'(r)\left({d t \over
d\tau}\right)^2-r\left({d \phi \over d\tau}\right)^2=0, \qquad {d^2
\theta \over d\tau^2}=0,\qquad {d^2 \phi \over d\tau^2}=0,\end{equation} and for
the geodesic deviation \begin{eqnarray}\label{ged11} && {d^2 \xi^1 \over d\tau^2}+w(r)w'(r) {dt \over d\tau}{d
\xi^0 \over d\tau}-2r w(r) {d \phi \over d\tau}{d \xi^3 \over
d\tau}+\left[{1 \over 2}\left(w'^2(r)+w(r) w''(r)
\right)\left({dt \over d\tau}\right)^2-\left(w(r)+rw'(r)
\right) \left({d\phi \over d\tau}\right)^2 \right]\xi^1=0, \nonumber\\
&&  {d^2 \xi^0 \over
d\tau^2}+{w'(r) \over w(r)}{dt \over d\tau}{d \zeta^1 \over d\tau}=0,\qquad {d^2 \xi^2 \over d\tau^2}+\left({d\phi \over d\tau}\right)^2
\xi^2=0, \qquad \qquad  {d^2 \xi^3 \over d\tau^2}+{2 \over r}{d\phi \over d\tau} {d
\xi^1 \over d\tau}=0, \end{eqnarray} where $w(r)$ is defined by the metric (\ref{me}) or (\ref{me1}),
$w'(r)=\displaystyle{dw(r) \over dr}$. Using
the circular orbit
\begin{equation} \theta={\pi \over 2}, \qquad
{d\theta \over d\tau}=0, \qquad {d r \over d\tau}=0,
\end{equation}
we get
\begin{equation}
 \left({d\phi \over d\tau}\right)^2={w'(r)
\over r[2w(r)-rw'(r)]}, \qquad \left({dt \over
d\tau}\right)^2={2 \over 2w(r)-rw'(r)}. \end{equation}

Eqs.~(\ref{ged11}) can be rewritten as
\begin{eqnarray} \label{ged22} &&  {d^2 \xi^1 \over d\phi^2}+w(r)w'(r) {dt \over
d\phi}{d \xi^0 \over d\phi}-2r w(r) {d \xi^3 \over
d\phi} +\left[{1 \over 2}\left[w'^2(r)+w(r) w''(r)
\right]\left({dt \over d\phi}\right)^2-\left[w(r)+rw'(r)
\right]  \right]\zeta^1=0, \nonumber\\
&&{d^2 \xi^2 \over d\phi^2}+\xi^2=0, \qquad {d^2 \xi^0 \over d\phi^2}+{w'(r) \over
w(r)}{dt \over d\phi}{d \xi^1 \over d\phi}=0,\qquad {d^2 \xi^3 \over d\phi^2}+{2 \over r} {d \xi^1 \over
d\phi}=0. \end{eqnarray}
The second equation  of (\ref{ged22}) corresponds to a simple harmonic motion, which  means that there is stability on the plane $\theta=\pi/2$. Assuming  the remaining equations of (\ref{ged22}) have solutions of the form
\begin{equation} \label{ged33}
\xi^0 = \zeta_1 e^{i \sigma \phi}, \qquad \xi^1= \zeta_2e^{i \sigma
\phi}, \qquad and \qquad \xi^3 = \zeta_3 e^{i \sigma \phi},
\end{equation}
where $\zeta_1, \zeta_2$ and $\zeta_3$ are constant and   $\phi$ is an unknown variable. Using Eq. (\ref{ged33}) into
(\ref{ged22}), one can get the stability condition for a static spherically symmetric charged black hole in the form
\begin{equation} \label{con1}  \displaystyle\frac{3ww'-\sigma^2w'-2rw'^2+rww''}{w'}>0.
\end{equation}
 Eq. (\ref{con1}) for solutions (\ref{me}) and (\ref{me1}) becomes
\begin{equation} \label{stab1}
\sigma^2=\frac{2r^2+\beta r^3+18{\cal K}^2\beta r+48\beta^2{\cal K}^4+4r^4\beta\Lambda[5r+4r^2\beta+24\beta{\cal K}^2]}{2\beta r^2(r+6\beta {\cal K}^2+4r^4\beta\Lambda)}>0.\end{equation}

\begin{figure}
\centering
\subfigure[~Stability of  (\ref{stab1}) when $\Lambda=0$ ]{\label{fig:11a}\includegraphics[scale=0.4]{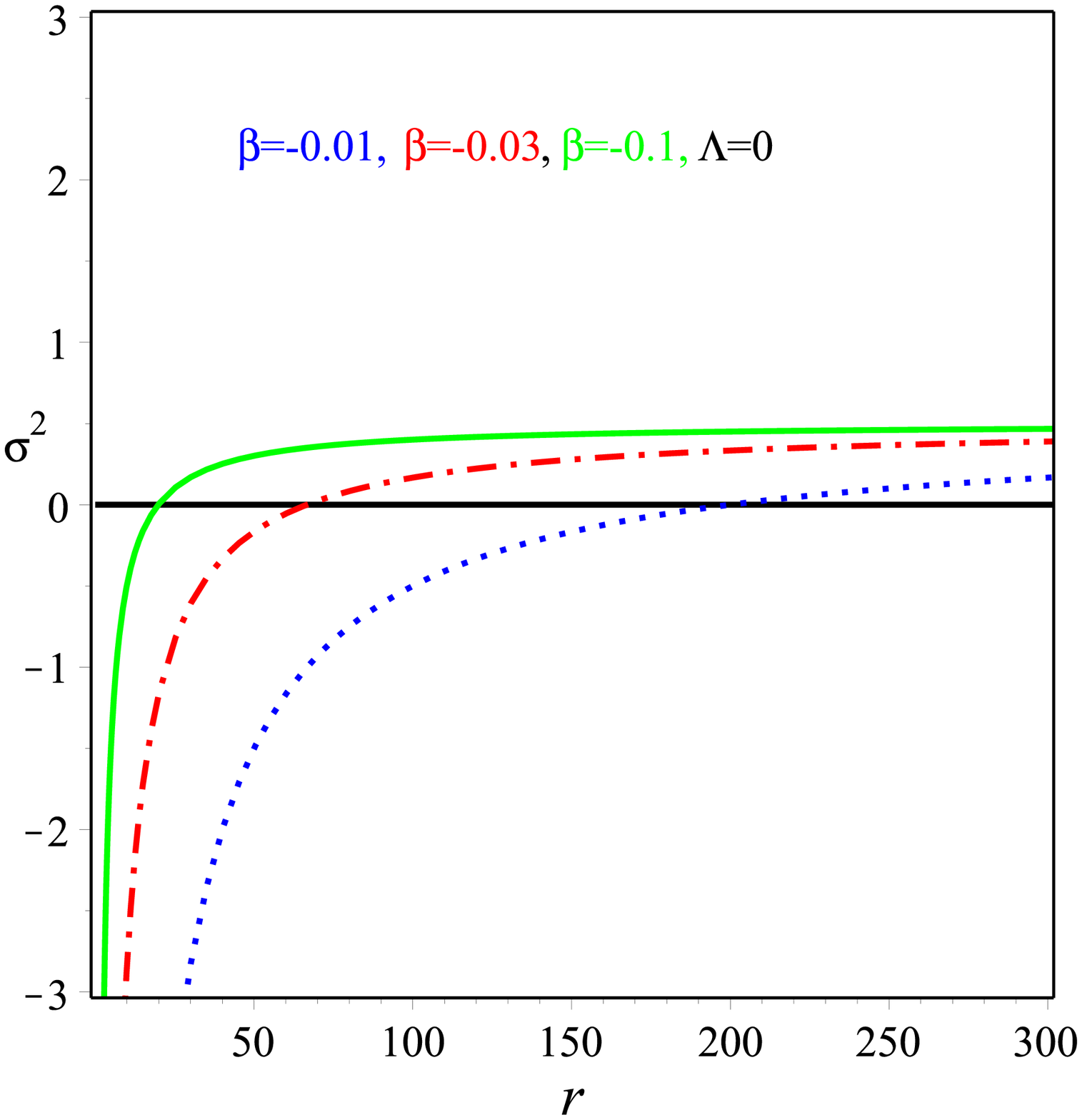}}\hspace{0.2cm}
\subfigure[~Stability of (\ref{stab1}) when $\Lambda\neq0$]{\label{fig:11b}\includegraphics[scale=0.4]{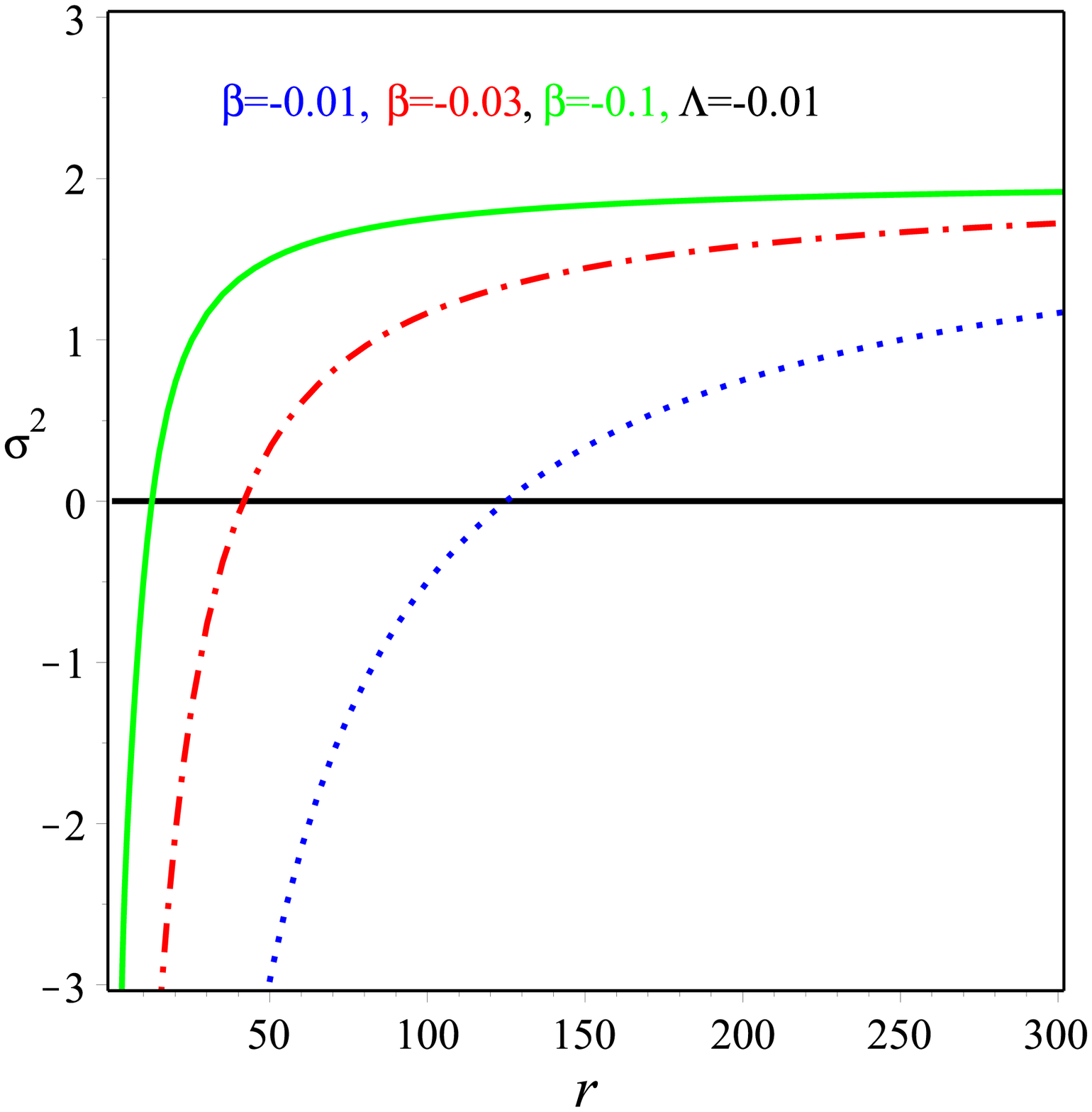}}
\caption{ {Schematic plot of Eq. (\ref{stab1}), namely $\sigma^2$  versus the coordinate $r$.}}
\label{Fig:11}
\end{figure}

Fig. \ref{Fig:11} is a plot of Eq. (\ref{stab1}) for particular values of the models. It exhibits the regions where the black holes are stable and the regions where there is no possible stability.

\subsection{Black hole stability analysis using geodesic deviation in Einstein's frame}\label{S6336}
 Introducing (\ref{ge}) and (\ref{ged})  into (\ref{conf}), we get for the geodesic equations
 \begin{equation}
{d^2 t \over d\tau^2}=0, \qquad {1 \over 2} w'_1(\bar{r})\left({d t \over
d\tau}\right)^2-\bar{r}\left({d \phi \over d\tau}\right)^2=0, \qquad {d^2
\theta \over d\tau^2}=0,\qquad {d^2 \phi \over d\tau^2}=0,
\end{equation}
and for
the geodesic deviation we have
\begin{eqnarray}\label{ged1} && {d^2 \xi^1 \over d\tau^2}+w(\bar{r})w'_1(\bar{r}) {dt \over d\tau}{d
\xi^0 \over d\tau}-2\bar{r} w(\bar{r}) {d \phi \over d\tau}{d \xi^3 \over
d\tau}+\left[{1 \over 2}\left[w'(\bar{r})w'_1(\bar{r})+w(\bar{r}) w''_1(\bar{r})
\right]\left({dt \over d\tau}\right)^2-\left[w(\bar{r})+\bar{r}w'(\bar{r})
\right] \left({d\phi \over d\tau}\right)^2 \right]\xi^1=0, \nonumber\\
&&  {d^2 \xi^0 \over
d\tau^2}+{w'_1(\bar{r}) \over w_1(\bar{r})}{dt \over d\tau}{d \zeta^1 \over d\tau}=0,\qquad {d^2 \xi^2 \over d\tau^2}+\left({d\phi \over d\tau}\right)^2
\xi^2=0, \qquad \qquad  {d^2 \xi^3 \over d\tau^2}+{2 \over \bar{r}}{d\phi \over d\tau} {d
\xi^1 \over d\tau}=0, \end{eqnarray}
where $w(\bar{r})$ is defined by the metric (\ref{conf})
$w'(\bar{r})=\displaystyle{dw(\bar{r}) \over d\bar{r}}$ and $w'_1(\bar{r})=\displaystyle{dw_1(\bar{r}) \over d\bar{r}}$. Using
the circular orbit  \begin{equation} \theta={\pi \over 2}, \qquad
{d\theta \over d\tau}=0, \qquad {d \bar{r} \over d\tau}=0,
\end{equation}
we get
\begin{equation}
 \left({d\phi \over d\tau}\right)^2={w'_1(\bar{r})
\over \bar{r}[2w(\bar{r})-\bar{r}w'(\bar{r})]}, \qquad \left({dt \over
d\tau}\right)^2={2 \over 2w(\bar{r})-w'(\bar{r})}. \end{equation}

\begin{figure}
\centering
\subfigure[~Stability of (\ref{ged2}) when $\Lambda=0$ ]{\label{fig:12a}\includegraphics[scale=0.4]{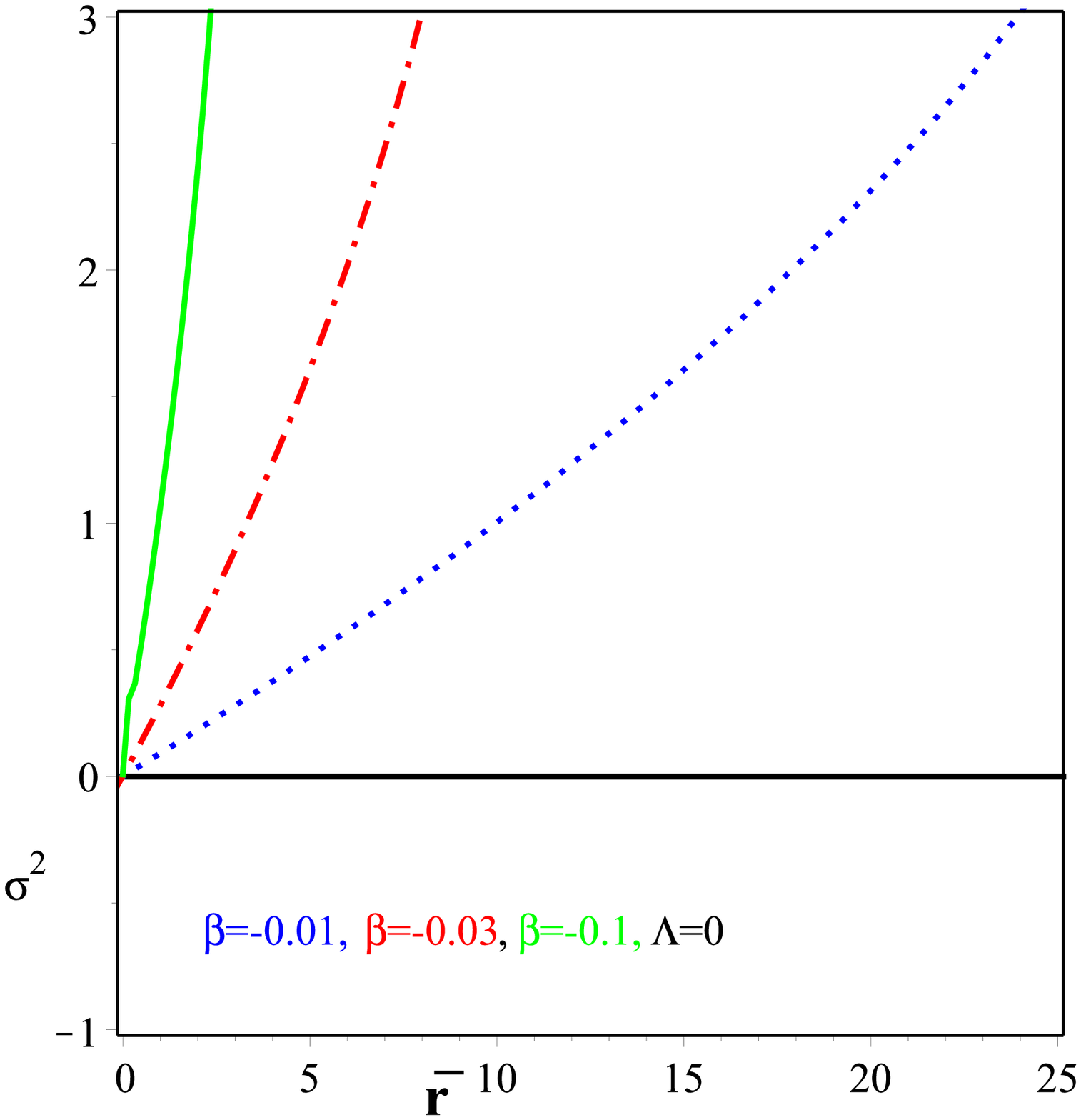}}\hspace{0.2cm}
\subfigure[~Stability of (\ref{ged2}) when $\Lambda\neq0$]{\label{fig:12b}\includegraphics[scale=0.4]{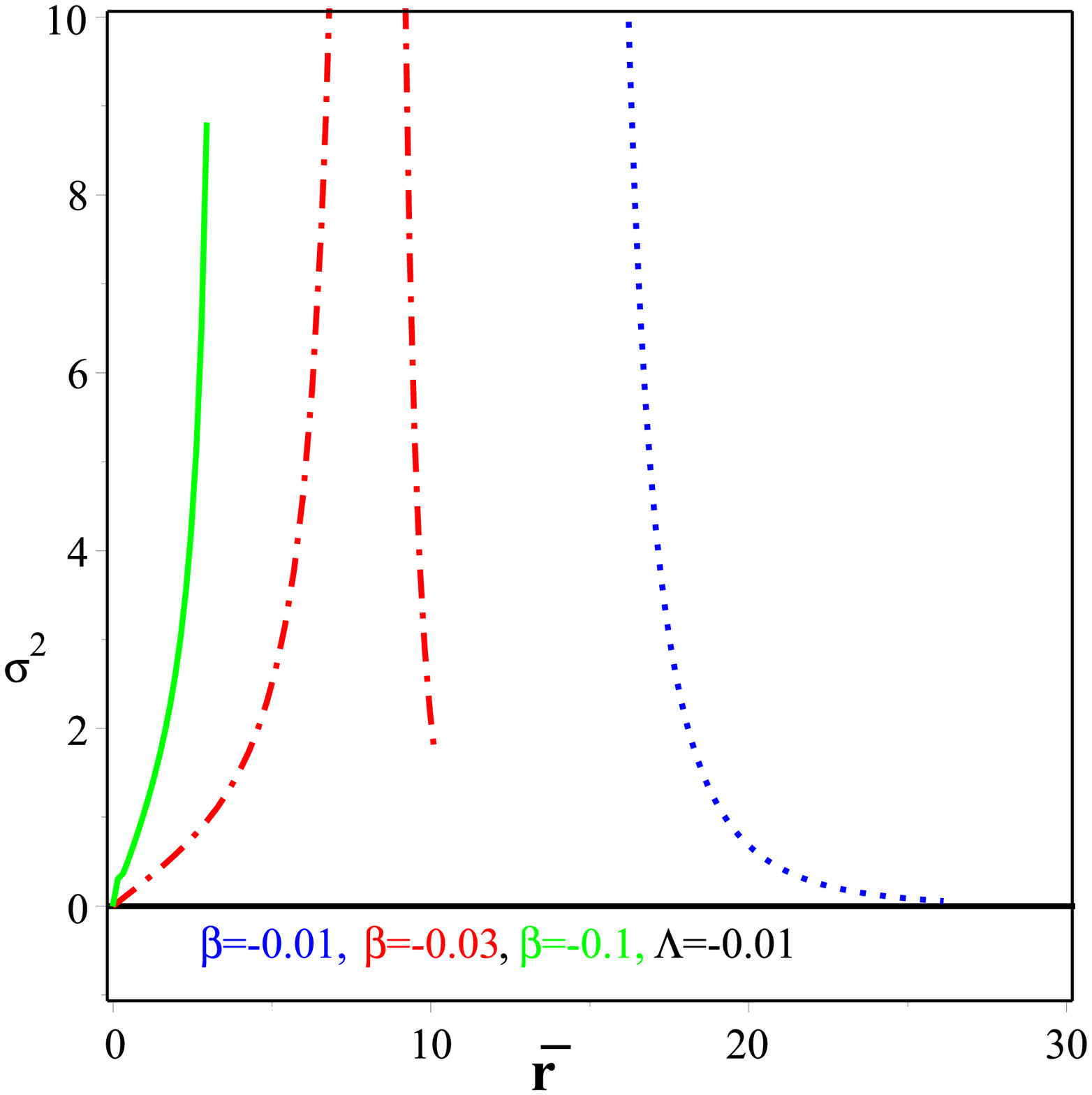}}
\caption{ {Schematic plot of Eq. (\ref{ged2}), namely $\sigma^2$  versus the coordinate $r$.}}
\label{Fig:12}
\end{figure}

Eqs. (\ref{ged1}) can be rewritten as
\begin{eqnarray} \label{ged2} &&  {d^2 \xi^1 \over d\phi^2}+w(\bar{r})w'_1(\bar{r}) {dt \over
d\phi}{d \xi^0 \over d\phi}-2\bar{r} w'(\bar{r}) {d \xi^3 \over
d\phi} +\left[{1 \over 2}\left[w'(\bar{r})w'_1(\bar{r})+w(\bar{r}) w''_1(\bar{r})
\right]\left({dt \over d\phi}\right)^2-\left[w(\bar{r})+\bar{r}w'(\bar{r})
\right]  \right]\zeta^1=0, \nonumber\\
&&{d^2 \xi^2 \over d\phi^2}+\xi^2=0, \qquad {d^2 \xi^0 \over d\phi^2}+{w'(\bar{r}) \over
w(\bar{r})}{dt \over d\phi}{d \xi^1 \over d\phi}=0,\qquad {d^2 \xi^3 \over d\phi^2}+{2 \over \bar{r}} {d \xi^1 \over
d\phi}=0. \end{eqnarray}
Eqs. (\ref{ged2}) corresponds to simple harmonic motion what  means we have a stability at the plane $\theta=\pi/2$. The remanning equations of (\ref{ged2}) admit  the following solutions
\begin{equation} \label{ged3}
\xi^0 = \zeta_1 e^{i \omega \phi}, \qquad \xi^1= \zeta_2e^{i \omega
\phi}, \qquad and \qquad \xi^3 = \zeta_3 e^{i \omega \phi},\end{equation}
where $\zeta_1, \zeta_2$ and $\zeta_3$ are constants and  $\phi$ is unknown. Using Eq. (\ref{ged3}) into
(\ref{ged2}) one gets  a quite lengthy expression, which is depicted in Fig.~\ref{Fig:12}. This plot shows that black holes in the Einstein frame have always some non-void stability region.

\subsection{Causal structure of the solutions}
We shall now discuss the causal structure of the space-time (\ref{met5}). For this purpose, we start from the metric
\begin{equation}
\label{GBiv}
ds^2 = - \e^{2\nu (r)} dt^2 + \e^{-2\nu (r)} dr^2
+ r^2 \sum_{i,j=1}^{2} \tilde g_{ij} dx^i dx^j\, ,
\end{equation}
with
\begin{equation}
\label{GBH1M}
\e^{2\nu} = C^2 - \frac{M}{r} \, , \quad C>0 \, ,
\end{equation}
 the metric of the unit sphere being $\tilde g_{ij}$, and
we consider the region where $r \gg M$.
Then, the metric in (\ref{GBiv}) reduces to
\begin{equation}
\label{de1}
ds_\mathrm{as}^2 = - C^2 dt^2 + \frac{dr^2}{C^2}
+ r^2 \sum_{i,j=1}^{2} \tilde g_{ij} dx^i dx^j\, .
\end{equation}
Redefining,
\begin{equation}
\label{de2}
t=\frac{\tilde t}{C}\, , \quad r=C\tilde r\, ,
\end{equation}
we find
\begin{equation}
\label{de3}
ds_\mathrm{as}^2 = - d{\tilde t}^2 + d{\tilde r}^2
+ C^2 {\tilde r}^2 \sum_{i,j=1}^{2} \tilde g_{ij} dx^i dx^j\, ,
\end{equation}
which is not Lorentz invariant unless $C=1$.
In order to clarify the situation, we choose $\tilde g_{ij}$ as
\begin{equation}
\label{de4}
\sum_{i,j=1}^{2} \tilde g_{ij} dx^i dx^j = d\theta^2 + \sin^2 \theta d\phi^2 \, ,
\end{equation}
with $0\leq \theta \leq \pi$ and $0\leq \phi < 2\pi$, and
we consider the hypersurface with $\theta=\frac{\pi}{2}$.
Then, the metric reads
\begin{equation}
\label{de5}
ds_\mathrm{hyp}^2 = - d{\tilde t}^2 + d{\tilde r}^2
+ C^2 d\phi^2 \, .
\end{equation}
If we redefine,
\begin{equation}
\label{de6}
\phi = C^{-1} \tilde\phi\, ,
\end{equation}
the metric acquires the following form
\begin{equation}
\label{de7}
ds_\mathrm{hyp}^2 = - d{\tilde t}^2 + d{\tilde r}^2  + d{\tilde\phi}^2 \, ,
\end{equation}
which is nothing but the metric of flat three-dimensional space-time.
We should note, however, that $0\leq \tilde\phi \leq 2C\pi$, and therefore if $C<1$, a deficit angle appears, while if $C>1$ a surplus angle shows up
(see Fig.~\ref{Fig1} for the case $C=\frac{3}{4}<1$ and Fig.~\ref{Fig2} for the case $C=\frac{5}{4}>1$).
For $C<1$ the light emitted from a point reaches  another point in two orbits of the light trajectory (see Fig.~\ref{Fig3}) and, therefore,  multiple light-cone surfaces are formed.
 On the contrary, in the other case the light ray emitted from a point $\tilde \phi=\pi$ and $\tilde r=r_0$
($r_0$ is a constant) does not reach the region $2\pi< \tilde\phi < 2C\pi$
(see Fig.~\ref{Fig4}) and, therefore, the light-cone surface has a boundary.
In such space-time, one cannot separate time-like regions from space-like
ones and all kind of problems with causality may show up.

\begin{figure}[h]
\centering

\unitlength=0.5mm
\begin{picture}(120,120)
\thinlines

\put(10,10){\line(1,0){100}}
\put(110,10){\line(0,1){100}}
\put(10,10){\line(0,1){50}}
\put(60,60){\vector(-1,0){50}}
\put(60,60){\vector(0,1){50}}
\put(60,110){\line(1,0){50}}


\thicklines


\end{picture}

\caption{Structure of the spatial part for $C=\frac{3}{4}<1$.
We identify two arrows and glue the space there. }\label{Fig1}
\end{figure}
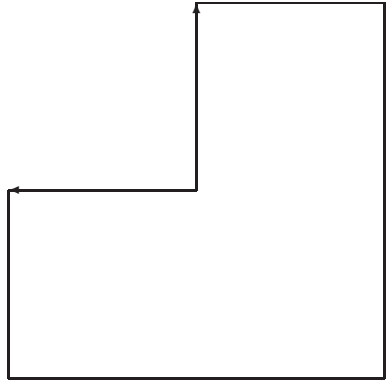

\begin{figure}[h]
\centering

\unitlength=0.5mm
\begin{picture}(120,120)
\thinlines

\put(10,10){\line(1,0){100}}
\put(110,10){\line(0,1){100}}
\put(10,10){\line(0,1){98}}
\put(10,108){\line(1,0){49}}
\put(60,110){\line(1,0){50}}

\put(106,60){\vector(1,0){2}}
\put(60,60){\vector(0,1){50}}

\qbezier[40](60,108)(84,108)(108,108)
\qbezier[40](108,108)(108,84)(108,60)
\qbezier[60](60,60)(84,60)(108,60)


\thicklines


\end{picture}

\caption{Structure of the spatial part for $C=\frac{5}{4}>1$.
We identify two arrow and glue the space there. }\label{Fig2}
\end{figure}
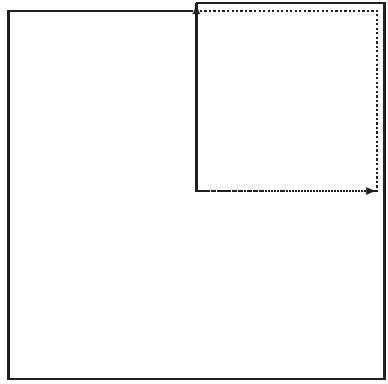

\begin{figure}[h]
\centering

\unitlength=0.5mm
\begin{picture}(120,120)
\thinlines

\put(10,10){\line(1,0){100}}
\put(110,10){\line(0,1){100}}
\put(10,10){\line(0,1){50}}
\put(60,60){\vector(-1,0){50}}
\put(60,60){\vector(0,1){50}}
\put(60,110){\line(1,0){50}}

\put(60.5,30){\makebox(0,0){$A$}}
\put(35.5,65){\makebox(0,0){$B$}}
\put(65.5,85){\makebox(0,0){$B'$}}

\thicklines

\put(60.5,35){\vector(0,1){50}}
\put(60.5,35){\vector(-1,1){25}}

\end{picture}

\caption{The point $B'$ is identified with the point $B$.
When $C=\frac{3}{4}<1$, the light emitted at  point $A$ may
reach point $B=B'$ along two different paths. }\label{Fig3}
\end{figure}
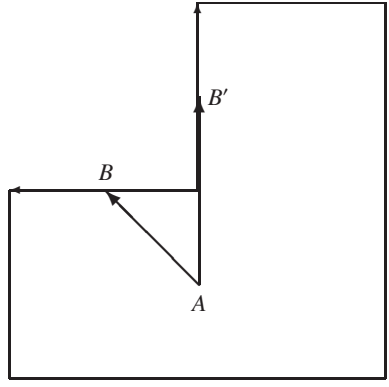

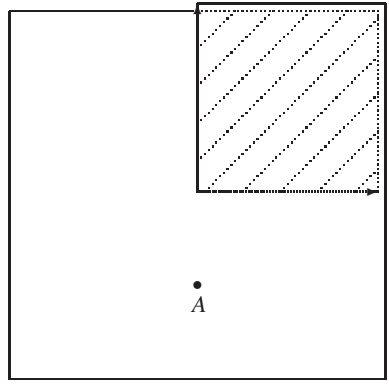
\begin{figure}[h]
\centering

\unitlength=0.5mm
\begin{picture}(120,120)
\thinlines

\put(10,10){\line(1,0){100}}
\put(110,10){\line(0,1){100}}
\put(10,10){\line(0,1){98}}
\put(10,108){\line(1,0){49}}
\put(60,110){\line(1,0){50}}

\put(106,60){\vector(1,0){2}}
\put(60,60){\vector(0,1){50}}

\qbezier[40](60,108)(84,108)(108,108)
\qbezier[40](108,108)(108,84)(108,60)
\qbezier[60](60,60)(84,60)(108,60)

\qbezier[10](60,98)(65,103)(70,108)
\qbezier[20](60,88)(70,98)(80,108)
\qbezier[30](60,78)(75,93)(90,108)
\qbezier[40](60,68)(80,88)(100,108)
\qbezier[46](62,60)(85,83)(108,106)
\qbezier[36](72,60)(90,78)(108,96)
\qbezier[26](82,60)(95,73)(108,86)
\qbezier[16](92,60)(100,68)(108,76)
\qbezier[6](102,60)(105,63)(108,66)

\put(60,35){\circle*{2}}
\put(60.5,30){\makebox(0,0){$A$}}

\thicklines


\end{picture}

\caption{When $C=\frac{5}{4}>1$, the light emitted at  point $A$
($\tilde \phi=\pi$ and $\tilde r=r_0$, $r_0$ is a constant)
cannot reach the shaded region ($2\pi< \tilde\phi < \frac{5}{2}\pi = 2C\pi$). }
\label{Fig4}
\end{figure}

\section{Alternative black hole description from a generalized fluid model}  \label{S66666}

\subsection{Relation between the space-time geometry and an equation of state}

We will here consider the relation existing between the                   space-time geometry and an equation of state for General Relativity with
              a    cosmological fluid. We start from the  space-time metric (\ref{GBiv}), from where we have

\begin{align}
\label{SEoS1}
R_{tt}=& \frac{1}{2} \left( C + \frac{r^2}{l^2} - \frac{M}{r} + \frac{q^2}{r^2} \right)
\left( \frac{6}{l^2} + \frac{2q^2}{r^4} \right) \, , \nn
R_{rr}=& - \frac{1}{2} \left( C + \frac{r^2}{l^2} - \frac{M}{r} + \frac{q^2}{r^2} \right)^{-1}
\left( \frac{6}{l^2} + \frac{2q^2}{r^4} \right) \, , \nn
R_{ij}=& \left( 1 - C - \frac{3r^2}{l^2} + \frac{q^2}{r^2} \right) \tilde g_{ij}\, , \nn
R=& - \frac{12}{l^2} - \frac{2\left( C - 1 \right)}{r^2} \, .
\end{align}
Using now the Einstein equation
\begin{equation}
\label{EinsteinEq}
R_{\mu\nu} - \frac{1}{2}g_{\mu\nu} R = \kappa^2 T_{\mu\nu} \, ,
\end{equation}
we find
\begin{align}
\label{SEoS2}
\kappa^2 T_{tt} =& - \kappa^2 T_{rr} =
\left( C + \frac{r^2}{l^2} - \frac{M}{r} + \frac{q^2}{r^2} \right)
\left( - \frac{3}{l^2}  - \frac{C - 1}{r^2} + \frac{q^2}{r^4} \right) \, , \nn
\kappa^2 T_{ij} =& \left( \frac{3r^2}{l^2} + \frac{q^2}{r^2} \right) \tilde g_{ij}
\end{align}
We now define the energy density $\rho$, the pressure in the radial direction $p_r$, and the pressure in the angular direction $p_a$, as
\begin{equation}
\label{SEoS3}
T_{tt}= - g_{tt} \rho\, , \quad T_{rr}= g_{rr} p_r \, , \quad T_{ij} = g_{ij} p_a \, ,
\end{equation}
with the result
\begin{equation}
\label{SEoS4}
\kappa^2 \rho = - \left( - \frac{3}{l^2}  - \frac{C - 1}{r^2} + \frac{q^2}{r^4} \right) \, , \quad
\kappa^2 p_r = - \left( - \frac{3}{l^2}  - \frac{C - 1}{r^2} + \frac{q^2}{r^4} \right) \, , \quad
\kappa^2 p_a = \frac{3}{l^2} + \frac{q^2}{r^4} \, ,
\end{equation}
which yields the following equation of state (EoS) for the cosmic fluid
\begin{equation}
\label{SEoS5}
\rho = p_r = - p_a + \frac{6}{\kappa^2 l^2}
+ \frac{C - 1}{\kappa^2 q} \sqrt{ \kappa^2 p_a - \frac{3}{l^2} }\, .
\end{equation}
In particular, when $\frac{1}{l}=0$, we find
\begin{equation}
\label{SEoS6}
\rho = p_r = - p_a + \frac{C - 1}{\kappa q} \sqrt{ p_a }\, .
\end{equation}
Eq.~(\ref{SEoS5}) tells us that $p_a\geq \frac{3}{\kappa^2 l^2}$, and so the quantity inside the square root is positive.
In particular, in the case $\frac{1}{l}=0$, as in (\ref{SEoS6}), we find $p_a\geq 0$.
And, in order that $\rho\geq 0$, we get $p_a\leq \frac{\left( C - 1\right)^2 }{\kappa^2 q^2}$.
Then, it follows that
\begin{equation}
\label{SEoS7}
0\leq p_a \leq \frac{\left( C - 1\right)^2 }{\kappa^2 q^2} \, .
\end{equation}
In the case $\frac{1}{l}\neq 0$, corresponding to Eq.~(\ref{SEoS5}), the restriction associated to (\ref{SEoS6}) becomes somehow involved, as follows
\begin{equation}
\label{SEoS8}
\frac{3}{\kappa^2 l^2} \leq p_a \leq \frac{1}{2} \left( \frac{12}{\kappa^2 l^2}
+ \frac{\left( C-1 \right)^2}{\kappa^2 q^2}
+ \sqrt{ \frac{12\left( C-1 \right)^2}{\kappa^4 l^2 q^2}
+ \frac{\left( C-1 \right)^4}{\kappa^4 q^4}} \right) \, .
\end{equation}
For $\frac{1}{l}=0$ in (\ref{SEoS6}), the pressures $p_r$ and $p_a$ should be positive if we assume that the energy density $\rho$ is positive.
We should note, however, that when $\frac{1}{l^2}<0$ in (\ref{SEoS5}), $p_a$ can be negative, as indeed found from (\ref{SEoS8}).
Therefore, this fluid can act as dark energy, what is indeed clear from the assumption (\ref{GBiv}), where the metric behaves as the de Sitter space-time for large $r$, when $\frac{1}{l^2}<0$.
\[
p_a^2 - \frac{12}{\kappa^2 l^2} + \frac{36}{\kappa^2 l^2} p_a
= \frac{\left( C-1 \right)^2}{\kappa^4 q^2} \left( \kappa^2 p_a - \frac{3}{l^2} \right)
\]
\begin{equation}
\label{SEoS8}
\frac{3}{\kappa^2 l^2} \leq p_a \leq \frac{1}{2} \left( \frac{12}{\kappa^2 l^2}
+ \frac{\left( C-1 \right)^2}{\kappa^2 q^2}
+ \sqrt{ \left( \frac{12}{\kappa^2 l^2}
+ \frac{\left( C-1 \right)^2}{\kappa^2 q^2} \right)^2
 - \frac{144}{\kappa^4 l^4}
 - \frac{12\left( C-1 \right)^2}{\kappa^4 l^2 q^2}} \right) \, .\end{equation}
 In summary, we have here derived the same BH solution as in usual general relativity with a cosmological fluid, which may be intrepreted as kin ad of dark energy. This is a clear indication of the universality of the BH
solution under discussion in this paper.

\section{Discussion and conclusions}\label{S77}

We have obtained, in this paper, a genuinely new type of charged black holes, with electric and magnetic charges, in the context of a particular class of $f(R)$ modified gravity. We have provided a detailed description of their physical properties, including their stability and causal structure, both in the Jordan and in the Einstein frames. Being more specific, we have worked with the following forms for $f(R)$, namely $f(R)=R+2\beta \sqrt{R}$ and $f(R)=R+2\beta\sqrt{R-8\Lambda}$,  to produce flat and AdS/dS space-times, respectively, and solved the field equations of $f(R)$ for a spherically symmetric space-time in which $g_{tt}=\frac{1}{g_{rr}}$\footnote{The reason for using a spherically symmetric space-time in which $g_{tt}=\frac{1}{g_{rr}}$ was simply to make the process of solving the $f(R)$ field equations more accessible, but variants of the same method could have been employed in less symmetric cases and more general situations.}. We have solved the resulting field equations in an exact way and derived black holes, which are characterized by three parameters: the mass, which depends on the dimensional parameter $\beta$, bound to have a negative value, and the electric and magnetic charges.

The Ricci scalar of the black holes here found is non-trivial. It has the form $R=\frac{1}{r^2}$, for the case of flat space-time, and $R=\frac{1}{r^2}+8\Lambda$ for the AdS/dS space-times. A most remarkable result is that these black holes cannot be reduced to the ordinary ones appearing in Einstein's GR; in other words, they are genuinely new black holes of the modified $f(R)$ gravities. We have calculated the scalar invariants of these solutions and shown that  $\beta\neq 0$. The calculations involving the scalar fields have shown that one gets a true singularity at $r=0$. Using conformal transformation, we got charged black hole solutions in the realm of the Einstein frame. An interesting feature of the black holes obtained in this frame is the fact that $g_{tt}\neq \frac{1}{g_{rr}}$, what does not happen for the corresponding black holes in the Jordan frame.  However, in spite of the fact that the black holes have different $g_{tt}$ and $g^{rr}$ components for the metric in the Einstein frame, they have coinciding  Killing  and event horizons.

It is well know that the Jordan and the Einstein frames are mathematically equivalent. To check if their corresponding associated physics are equivalent, too, we have calculated some thermodynamical quantities for the above black holes, respectively obtained in one and in the other frame. A detailed discussion has shown that the physics associated with the entropy, quasi-local energy, and Gibbs free energy, in both frames, turn out to be fully equivalent. However, the physics associated to the Hawking temperature is not the same in both frames: the temperature in the Jordan frame is always positive, contrary to what happens in the Einstein frame, which can lead to negative values of yhe same. This may serve as an indication that the physics of the two frames are not equivalent, at least concerning this important quantity, the black hole temperature. An intriguing conjecture that has come to our minds is the following: could this possibly be related to the loss-of-information paradox?

Going more deeply into the black hole properties, we have studied their stability using linear perturbations. Our calculations show that the radial propagation speed always equals one, in both frames, which means that the constructed black holes are stable. In addition, we have used the procedure of geodesic deviation to study the stability of the black holes, both  in the Jordan and in the Einstein frames, and derived in each case the stability condition.    Finally, we have also studied the causal structure of our novel black holes. We have shown that, in general, they are not invariant under Lorentz transformations. Moreover, we have identified that there is a (positive or negative) deficit angle associated with them. It goes without saying that the black holes obtained in this work need still to be analyzed in more depth, in order to unveil all of their physical properties, a job we hope to undertake elsewhere. Furthermore, an extension of this study to less symmetric backgrounds and to a more general form of $f(R)$ is pending.

We now consider the possibility that the black hole corresponding to
the solution (\ref{met5})
can be found by any observation.
The analysis in Section IXB tell that the solution (\ref{met5})
corresponds to $C=\frac{1}{\sqrt{2}}$ in  (\ref{GBiv})
with  (\ref{GBH1M}).
Because $C=\frac{1}{\sqrt{2}}<1$, the solution makes the deficit angle.
Then Figure \ref{Fig3}
tells that the black hole generates strong gravitational lensing effects.
In the usual black hole, the lensing effects occur only in the region near the black hole but
for the geometry expressed by the metric  (\ref{met5})
the effects occur in a rather large region, say interstellar region, around the black hole.
Therefore the big ring much greater than the standard Einstein ring or double images separated
in a large angle could be observed as in the observation of the standard weak lensing as in
\cite{Hildebrandt:2016iqg,Joudaki:2017zdt,Aghanim:2018oex}
in future.
\\
\\
\centerline{\bf{Appendix A}}
\centerline{\bf{The field equations of Ansatz (\ref{met}) without cosmological constant }}
Imposing the  Ansatz (\ref{met}) to Eqs. (\ref{f1}), (\ref{fe2}) and (\ref{f3}), after using Eq. (\ref{r1}), we get\footnote{Here in these calculations we  set $\Lambda=0$. }
$$
\zeta_t{}^t=\frac{1}{4r^{6}\sin^2\theta \sqrt{{\cal R}^5}}\Big\{4r^2\sqrt{{\cal R}^5}\Big[r^2\sin^2\theta(rw'+wn_\theta{}^2-2ws'n_\theta+q'^2r^2-1+w\{1+s'^2\})+k_\theta{}^2+r^2wp'^2\Big]+\beta\sin^2\theta \Big[r^6w{\cal R}w''''+3r^6ww'''^2\nonumber\\
$$
$$
-r^3w'''(r^2w''[rw'-12w]+4r^2w'^2+2rw'[31w-1]-48w[1-w])+2r^6w''^3-4r^4w''^2(4-6rw'-15w)+2r^2w''(57r^2w'^2\nonumber\\
$$
$$
+14r[3w-5]w'-36w^2+20+16w)+200r^3w'^3+4r^2w'^2(96w-85)+8rw'[w-1][27w-23]+32[w-1]^2[2w-1]\Big]\Big\}=0,\eqno{({ A\cdot 1})}\nonumber\\
$$
$$
\zeta_t{}^\theta=\frac{2q'w(n_\theta-s')}{r^2}=0,\qquad  \zeta_t{}^\phi=\frac{2q'fp'}{r^2\sin^2\theta}=0,\qquad \zeta_r{}^\theta=\frac{2p'k_\theta}{r^4\sin^2\theta}=0,\qquad \zeta_r{}^\phi=\frac{2(n_\theta-s')k_\theta}{r^4\sin^2\theta}=0,\eqno{({ A\cdot 2})}\nonumber\\
$$
$$ \zeta_r{}^r=\frac{1}{4r^4\sin^2\theta\sqrt{{\cal R}^3}}\Big\{4\sqrt{{\cal R}^3}\Big[r^2\sin^2\theta(rw'-wn_\theta{}^2+2ws'n_\theta-ws'^2+[w-1+r^2q'^2])+k'^2
-wr^2p'^2\Big]+\beta \sin^2\theta \Big(r^2w'''[4w+rw']\nonumber\\
$$
$$
-2r^4w''^2+w''[4r^2(3+w)-16r^3w']-50r^2w'^2+4rw'(15-17w)-16(1-3w+2w^2)\Big)\Big\}=0,\eqno{({ A\cdot 3})}\nonumber\\
$$
$$
\zeta_\theta{}^t=2q'(n_\theta-s')=0,\qquad \zeta_\theta{}^r=\frac{2wk_\theta p'}{r^2\sin^2\theta},\qquad \zeta_\theta{}^\phi=\frac{2w(n_\theta-s')p'}{r^2\sin^2\theta},\qquad \zeta_\phi{}^t=2q'p'=0, \eqno{({ A\cdot 4})}\nonumber\\
$$
$$
 \zeta_\phi{}^r=\frac{2w(n_\theta-s')k_\theta}{r^2}=0,\qquad \zeta_\phi{}^\theta=\frac{2wp'(n_\theta-s')}{r^2}=0,\eqno{({ A\cdot 5})}\nonumber\\
$$
$$ \zeta_\theta{}^\theta=
\frac{1}{2r^{6}\sin^2\theta \sqrt{{\cal R}^5}}\Big\{2r^2\sqrt{{\cal R}^5}\Big[r^2\sin^2\theta\Big(\frac{r^2w''}{2}+rw'-wn_\theta{}^2+2wn_\theta s'-q'^2r^2-ws'^2\Big)-k_\theta{}^2+2l_\phi k_\theta+r^2wp'^2\Big]-\beta\sin^2\theta \Big[r^6w{\cal R}w''''\nonumber\\
$$
$$
+\frac{3r^6ww'''^2}{2}-r^3w'''(r^2w''[rw'-7w]+4r^2w'^2+2rw'[14w-1]-22w[1-w])+2r^6w''^3-2r^4w''^2\Big(5-9rw'-18w\Big)+2r^2w''(33r^2w'^2\nonumber\\
$$
$$
+r[27w-34]w'-18w^2+8+10w)]+104r^3w'^3+2r^2w'^2(81w-74)+4rw'[15w-16][w-1]-4(2-8w+10w^2-4w^3)\Big\}=0,\eqno{({ A\cdot 6})}\nonumber\\
$$
$$
\zeta_\phi{}^\phi=
\frac{1}{2r^{6}\sin^2\theta \sqrt{{\cal R}^5}}\Big\{2r^2\sqrt{{\cal R}^5}\Big[r^2\sin^2\theta\Big(\frac{r^2w''}{2}+rw'+wn_\theta{}^2-2wn_\theta s'-q'^2r^2+ws'^2\Big)-k_\theta{}^2-r^2wp'^2\Big]-\beta\sin^2\theta \Big[r^6w{\cal R}w''''\nonumber\\
$$
$$
+\frac{3r^6ww'''^2}{2}-r^3w'''(r^2w''[rw'-7w]+4r^2w'^2+2rw'[14w-1]-22w[1-w])+2r^6w''^3-2r^4w''^2\Big(5-9rw'-18w\Big)+2r^2w''(33r^2w'^2\nonumber\\
$$
$$
+r[27w-34]w'-18w^2+8+10w)]+104r^3w'^3+2r^2w'^2(81w-74)+4rw'[15w-16][w-1]-4(2-8w+10w^2-4w^3)\Big\}=0,\eqno{({ A\cdot 7})}\nonumber\\
$$
$$
\zeta =
\frac{1}{2r^{6}\sqrt{{\cal R}^5}}\Big\{2r^6\sqrt{{\cal R}^5}-3\beta \Big[r^6w{\cal R}w''''+\frac{3r^6ww'''^2}{2}-r^3w'''(r^2w''[rw'-6w]+4r^2w'^2+2rw'[31w-1]-24w[1-w])+2r^6w''^3\nonumber\\
$$
$$
-2r^4w''^2\Big(6-10rw'-17w\Big)+2r^2w''(41r^2w'^2+2r[16w-23]w'-16w^2+12+4w)+136r^3w'^3+2r^2w'^2(117w-106)\nonumber\\
$$
$$
+8rw'[15w-13][w-1]+16(2w-1)(w-1)^2\Big]\Big\}=0, \eqno{({ A\cdot 8})} \nonumber\\
$$
where\footnote{We set $w(r)=w$, $q(r)=q$, $n(\theta)=n$,  $s(r)=s$,  $p(r)=p$, and $k(\theta)=k$.} $q(r)$, $n(\theta)$, $s(r)$, $l(\phi)$, $p(r)$, and $k(\theta)$ are  the  gauge potentials, defined as
$$ \xi :=q(r)dt+ n(\theta)dr+s(r)d\theta+[p(r)+k(\theta)]d\phi. \eqno{({ A\cdot 9})} $$   For brevity, we put $w'=\frac{dw}{dr}$, $w''=\frac{d^2w}{dr^2}$, $w'''=\frac{d^3w}{dr^3}$, $w''''=\frac{d^4w}{dr^4}$, $q'=\frac{dq}{dr}$, $s'=\frac{ds}{dr}$, $p'=\frac{dm}{dr}$ $n_\theta=\frac{dn}{d\theta}$ and  $k_\theta=\frac{dk}{d\theta}$. We must note that, when the magnetic fields vanish, i.e. $n=s=p=k=0$, we get $\zeta_\theta{}^\theta=\zeta_\phi{}^\phi$, and in this case the field equation $({ A\cdot 1})$ $\sim$ $({ A\cdot 8})$ coincides with the one derived in \cite{Nashed:2019tuk}.\vspace{0.3cm}\\

\centerline{\bf{Appendix B}}
\centerline{\bf{The field equations of Ansatz (\ref{met}) with cosmological constant }}
Now
$$
 \zeta_t{}^t=\frac{1}{4r^{9}\sin^2\theta \sqrt{{\textrm R}^5}}\Big\{4r^5\sqrt{{\textrm R}^5}\Big[r^2\sin^2\theta(rw'+wn_\theta{}^2-2ws'n_\theta+q'^2r^2-1+2r^2\Lambda+w\{1+s'^2\})+k_\theta{}^2+r^2+wp'^2\Big]+\beta\sin^2\theta \Big[r^6w{\textrm R}w''''\nonumber\\
$$
$$
+3r^6ww'''^2-r^3w'''(r^2w''[rw'-12w]+4r^2w'^2+2rw'[31w-1+4r^2\Lambda]+48w[1-w-2r^2\Lambda])+2r^6w''^3-4r^4w''^2\Big(4-6rw'-15w\nonumber\\
$$
$$
-16r^2\Lambda\Big)+2r^2w''(57r^2w'^2+14r[3w-5+8r^2\Lambda]w'-36w^2+80[4r^2\Lambda-1]^2+16w[1+9r^2\Lambda])+200r^3w'^3+4r^2w'^2\Big(96w-85\nonumber\\
$$
$$
+324r^2\Lambda\Big)+8rw'\Big\{27w^2-23+
2w(47r^2\Lambda-50)+11w^3+5w^2(1-2r^2\Lambda)+w(2+24r^4\Lambda^2-11r^2\Lambda)+(4r^2\Lambda-1)^3\Big\}\Big]\Big\}=0,\eqno{({ B\cdot 1})}\nonumber\\
$$
$$
\zeta_t{}^\theta=\frac{2q'w(n_\theta-s')}{r^2}=0,\qquad  \zeta_t{}^\phi=2q'wp'=0,\qquad \zeta_r{}^\theta=2p'k_\theta=0,\qquad \zeta_r{}^\phi=\frac{2(n_\theta-s')k_\theta}{r^4\sin^2\theta}=0, \eqno{({ B\cdot 2})}\nonumber\\
$$
$$
 \zeta_r{}^r=\frac{1}{2r^7\sin^2\theta \sqrt{{\textrm R}^3}}\Big\{r \sqrt{{\textrm R}^3}\Big[r^2\sin^2\theta(rw'-wn_\theta{}^2+2ws'n_\theta-ws'^2+r^2q'+w-1+2r^2\Lambda)+k_\theta{}^2-r^2+wp'^2\Big]+\beta \sin^2\theta r^3\Big[r^3w'''\Big(4w\nonumber\\
$$
$$
+rw'\Big)-2r^4w''^2+4r^2w''(12r^2\Lambda-3-w+4rw')-50r^2w'^2-4rw'(56r^2\Lambda-15+17w)-32w^2+16w(3-16r^2\Lambda)\nonumber\\
$$
$$
 \hspace*{-12cm}+16(1-4r^2\Lambda)^2\Big]\Big\}=0, \eqno{({ B\cdot 3})}\nonumber\\
$$
$$
\zeta_\theta{}^t=2q'(n_\theta-s')=0,\qquad \zeta_\theta{}^r=\frac{2w(l_\phi-k_\theta)p'}{r^2\sin^2\theta}=0,\qquad \zeta_\theta{}^\phi=\frac{2w(n_\theta-s')p'}{r^2\sin^2\theta}=0,\qquad \zeta_\phi{}^t=2q'p'=0,   \eqno{({ B\cdot 4})} \nonumber\\
$$
$$
\zeta_\phi{}^r=\frac{2w(n_\theta-s')k_\theta}{r^2}=0,\qquad \qquad  \zeta_\phi{}^\theta=\frac{2wp'(n_\theta-s')}{r^2}=0,\eqno{({ B\cdot 5})}\nonumber\\
$$
$$
 \zeta_\theta{}^\theta=\frac{1}{2r^9\sin^2\theta \sqrt{{\textrm R}^5}}\Big\{4r^5 \sqrt{{\textrm R}^5}\Big[r^2\sin^2\theta(r^2w''+2rw'-2wn_\theta{}^2+4ws'n_\theta-2ws'^2-2r^2q'^2+4r^2\Lambda)-2k_\theta{}^2+2r^2+2wp'^2\Big]\nonumber\\
$$
$$
-\beta \sin^2\theta\Big[3r^6ww'''^2-4r^6ww''''{\textrm R}-4r^3w'''([rw'-7w]r^2w''+4r^2w'^2+2rw'[4r^2\Lambda-1+14w]+2w[11w+20r^2\Lambda-11])\nonumber\\
$$
$$
+8r^6w''^3+8r^4w''^2[9rw'+18w-5+24r^2\Lambda]+8r^2w''\Big(33r^2w'^2+rw'[32r^2\Lambda-34+27w]-18w^2+2w[5+33r^2\Lambda]\nonumber\\
$$
$$
+4[1-10r^2\Lambda
+24r^4\Lambda^2]\Big)+208r^3w'^3+8r^2w'^2[81w-74+328r^2\Lambda]+32rw'(15w^2+w[156r^2\Lambda-31]+8[2+82r^4\Lambda^2-19r^2\Lambda])\nonumber\\
$$
$$
+32w^3+32w^2[8r^2\Lambda-5]-64w(6r^2\Lambda-1-20r^4\Lambda^2)+32(1-4r^2\Lambda)^2(8r^2\Lambda-1)\Big]\Big\}=0, \eqno{({ B\cdot 6})}\nonumber\\
$$
$$
 \zeta_\phi{}^\phi=\frac{1}{2r^9\sin^2\theta \sqrt{{\textrm R}^5}}\Big\{4r^5 \sqrt{{\textrm R}^5}\Big[r^2\sin^2\theta(r^2w''+2rw'+2wn_\theta{}^2-4ws'n_\theta+2ws'^2-2r^2q'^2+4r^2\Lambda)-2k_\theta{}^2-2r^2+2wp'^2\Big]\nonumber\\
$$
$$
-\beta \sin^2\theta\Big[3r^6ww'''^2-4r^6ww''''{\textrm R}-4r^3w'''([rw'-7w]r^2w''+4r^2w'^2+2rw'[4r^2\Lambda-1+14w]+2w[11w+20r^2\Lambda-11])\nonumber\\
$$
$$
+8r^6w''^3+8r^4w''^2[9rw'+18w-5+24r^2\Lambda]+8r^2w''(33r^2w'^2+rw'[32r^2\Lambda-34+27w]-18w^2+2w[5+33r^2\Lambda]\nonumber\\
$$
$$
+4[1-10r^2\Lambda
+24r^4\Lambda^2])+208r^3w'^3+8r^2w'^2[81w-74+328r^2\Lambda]+32rw'(15w^2+w[156r^2\Lambda-31]+2[8+321r^4\Lambda^2-76r^2\Lambda])\nonumber\\
$$
$$
+32w^3+32w^2[8r^2\Lambda-5]-64w(6r^2\Lambda-1-20r^4\Lambda^2)+32(1-4r^2\Lambda)^2(8r^2\Lambda-1)\Big]\Big\}=0, \eqno{({ B\cdot 7})}\nonumber\\
$$
$$
 \zeta= \frac{1}{4r^6\sqrt{{\textrm R}^5}}\Big\{4r^6 \sqrt{{\textrm R}^7}+\beta \Big[6r^6{\textrm R}ww''''+9r^6ww'''^2-6r^3w'''([rw'-6w]r^2w''+4r^2w'^2+2rw'[4r^2\Lambda-1+16w]\nonumber\\
$$
$$
+24w[w-1+2r^2\Lambda])+12r^6w''^3+4r^4w''^2[30rw'+51w-18+80r^2\Lambda]+4r^2w''(123r^2w'^2+22rw'[296r^2\Lambda-69+48w]-48w^2\nonumber\\
$$
$$
+4w[3+74r^2\Lambda]+4[9-80r^2\Lambda
+176r^4\Lambda^2])+816r^3w'^3+4r^2w'^2[351w-318+1304r^2\Lambda]+16rw'(45w^2+w[344r^2\Lambda-84]+39\nonumber\\
$$
$$
+704r^4\Lambda^2-332r^2\Lambda)+192w^3+32w^2[34r^2\Lambda-15]-64w(37r^2\Lambda-6-88r^4\Lambda^2)+32(1-4r^2\Lambda)^2(16r^2\Lambda-3)\Big]\Big\}=0.\eqno{({ B\cdot 8})}\nonumber
$$
\section*{Acknowledgments}
EE and SDO have been partially supported by MINECO (Spain), Project FIS2016-76363-P, and by the CPAN Consolider Ingenio 2010 Project. SN by a MEXT KAKENHI Grant-in-Aid for Scientific Research on Innovative Areas ``Cosmic Acceleration'' (No. 15H05890).
%

\end{document}